\documentclass[journal]{IEEEtran}
\UseRawInputEncoding
\usepackage{xcolor}
\usepackage[all]{nowidow}
\usepackage{bm}
\usepackage[ruled,vlined]{algorithm2e}
\usepackage{paralist}
\usepackage{mathtools, cuted, amsmath, amsthm, amssymb}
\usepackage[english]{babel}
\usepackage{tikz}
\usepackage{pgfplots}
\usetikzlibrary{patterns}
\usepackage{nomencl}
\pgfplotsset{compat=1.10}
\usepgfplotslibrary{fillbetween}
\RequirePackage{ifthen}
\usepackage{enumitem}
\usepackage{bbm}
\usepackage{flushend}
\usepackage{forest}
\usetikzlibrary{shapes.gates.logic.US,trees,positioning,arrows}
\pgfplotsset{compat = newest}
\newtheorem{theorem}{Theorem}[section]

\newtheorem{lemma}[theorem]{Lemma}
\usepackage{fancybox}

\usepackage{subfigure}

\usetikzlibrary{decorations.text}
\usetikzlibrary{positioning,shapes,arrows}

\tikzstyle{myBlock}=[draw, fill=blue!20!white, minimum size=0.5in, node distance=1in, text width=1.5cm, align=center]
\tikzstyle{myPath} = [-{Latex[length=2mm,width=2mm]}, line width=0.2mm]

\usetikzlibrary{arrows.meta}

\makenomenclature

\usepackage{cite}
\usepackage{amsmath,amssymb,amsfonts}
\usepackage{algorithmic}
\usepackage{graphicx}
\usepackage{textcomp}
\def\BibTeX{{\rm B\kern-.05em{\sc i\kern-.025em b}\kern-.08em
T\kern-.1667em\lower.7ex\hbox{E}\kern-.125emX}}
\definecolor{light-gray}{gray}{0.95}
\newcommand*\circled[1]{\tikz[baseline=(char.base)]{%
        \node[shape=circle,draw,inner sep=1pt] (char) {#1};}}
\usepackage{tikz}
\usepackage{enumitem}
\begin{document}
%
\title{Cyber LOPA: An Integrated Approach for the Design of Dependable and Secure Cyber Physical Systems}
%
%
%

\author{Ashraf~Tantawy,~\IEEEmembership{Member,~IEEE,}
    Sherif~Abdelwahed,~\IEEEmembership{Senior~Member,~IEEE,}
    and~Abdelkarim~Erradi,~\IEEEmembership{Member,~IEEE}
\thanks{Ashraf Tantawy is with the School of Computer Science and Informatics, De Montfort University, Leicester LE1 9BH, UK. E-mail: ashraf.tantavy@dmu.ac.uk}
\thanks{Sherif Abdelwahed is with the Department of Electrical and Computer Engineering, Virginia Commonwealth University, Richmond, VA 23284, USA. E-mail: sabdelwahed @vcu.edu}
\thanks{Abdelkarim Erradi is with the Department of Computer Science and Engineering, Qatar University, Doha, Qatar. Email: erradi@qu.edu.qa}
}
{\let\newpage\relax\maketitle}
\begin{abstract}
Safety risk assessment is an essential process to ensure a dependable Cyber-Physical System  (CPS) design. Traditional risk assessment considers only physical failures. For modern CPS, failures caused by cyber attacks are on the rise. The focus of latest research effort is on safety-security lifecycle integration and the expansion of modeling formalisms for risk assessment to incorporate security failures. The interaction between safety and security lifecycles and its impact on the overall system design, as well as the reliability loss resulting from ignoring security failures are some of the overlooked research questions. This paper addresses these research questions by presenting a new safety design method named Cyber Layer Of Protection Analysis (CLOPA) that extends existing LOPA framework to include failures caused by cyber attacks.  The proposed method provides a rigorous mathematical formulation that expresses quantitatively the trade-off between designing a highly-reliable versus a highly-secure CPS. We further propose a co-design lifecycle process that integrates the safety and security risk assessment processes. We evaluate the proposed CLOPA approach and the integrated lifecycle on a practical case study of a process reactor controlled by an industrial control testbed, and provide a comparison between the proposed CLOPA and current LOPA risk assessment practice.
\end{abstract}


%
\IEEEpeerreviewmaketitle

\nomenclature[A]{CPS}{Cyber-Physical System}
\nomenclature[A]{CSTR}{Continuous Stirred Tank Reactor}
\nomenclature[A]{BPCS}{Basic Process Control System}
\nomenclature[A]{SIS}{Safety Instrumented System}
\nomenclature[A]{HMI}{Human Machine Interface}
\nomenclature[A]{DMZ}{DeMilitarized Zone}
\nomenclature[A]{HAZOP}{HAZard and OPerability}
\nomenclature[A]{SIF}{Safety Instrumented Function}
\nomenclature[A]{IPL}{Independent Protection Layer}
\nomenclature[A]{GUI}{Graphical User Interface}
\nomenclature[A]{P\&ID}{Process \& Instrumentation Diagram}
\nomenclature[A]{MITM}{Man In The Middle}
\nomenclature[A]{SSH}{Secure Socket Shell}
\nomenclature[A]{DoS}{Denial of Service}
\nomenclature[A]{PID}{Proportional Integral Derivative}
\nomenclature[A]{LOPA}{Layer Of Protection Analysis}
\nomenclature[A]{CLOPA}{Cyber Layer of Protection Analysis}
\nomenclature[A]{TMEL}{Target Mitigated Event Likelihood}
\nomenclature[A]{RRF}{Risk Reduction Factor}
\nomenclature[A]{SIL}{Safety Integrity Level}	
\printnomenclature[1.5cm]

\section{Introduction}
\label{sec:introduction}


%
%
%
%
\IEEEPARstart{A}{} cyber physical system (CPS) is an integration of a physical process with computation and networking required for physical system monitoring and control. The integration of process dynamics with those of computation and networking brings a plethora of engineering challenges. As the majority of CPS are deployed in mission-critical applications, the dependability and resilience to failures is a key design property for modern CPS.

To ensure a given CPS is dependable, a risk assessment is carried out both at design time and operation time. The risk assessment process highlights the system weaknesses and helps define the safety requirements that need to be met to achieve the target reliability measures. The classical approach to perform the risk assessment is to consider physical failures only. As state of the art CPS designs move to open source hardware and software, cyber attacks have become a source of failure that cannot be ignored.

Realizing the critical nature of CPS cyber attacks and their impact on the safety of people and environment, as well as the potential catastrophic financial losses, the research community developed several approaches to integrate security aspects into the safety risk assessment process. This integration has been done mainly by extending the reliability modeling formalism to incorporate security-related risks. One of the overlooked research questions is how safety and security interact with each other and how this interaction would impact the overall system design. Putting this research question in a different format: \textit{Is there a trade-off between designing a highly reliable and a highly secure system?} A related research question is: \textit{If we ignore the cyber security attacks in the design process, what is the impact on the overall system reliability? Is the reliability gain worth the complexity introduced by integrating security both at design and run-time?} A follow-up research question is: \textit{Under what conditions can we ignore security failures?}

In order to better understand the interaction between safety and security lifecycles in the system design process, we consider in this paper the safety risk assessment process and study the impact of overlooking failures caused by cyber attacks. We refer to such failures as security failures in the rest of the paper. By formally introducing the failures caused by attacks into the risk assessment process, we can define the reliability requirements for the cyber components of the system as a function of both the failure rate of physical components and the resilience to cyber attacks. This formal requirements specification enables us to understand the design trade-off between higher reliability of physical components vs higher resilience of cyber components, and the sensitivity of the overall system performance to both types of failures. In addition, we can gain insight into the interplay between safety and security and how to integrate both lifecycles during the design process. More specifically, we consider the Layer Of Protection Analysis (LOPA), a widely adopted risk assessment method that follows a hazard identification study, such as Hazard and Operability (HAZOP). LOPA is carried out to identify whether an additional Safety Instrumented System (SIS) is needed for specific hazardous scenarios to achieve the target risk level. As modern SIS is typically an embedded device, it has both physical and security failure modes. We mathematically derive the SIS design constraints in terms of both physical and security failure probabilities. Additionally, we propose an integrated safety-security design process that shows the flow of information between both lifecycles.

We can classify the research work on combining safety and security for CPS into two broad categories that try to answer the following research questions: \begin{inparaenum}[(1)] \item Given the independent safety and security lifecycles, what are the similarities/differences and how could the two lifecycles be aligned or unified? This research direction usually focuses on answering the question "what to do", rather than "how to do it", \item For a given CPS, how can we carry out risk assessment (qualitative/quantitative) that considers both physical failures and cyber attacks? Consequently, how can we unify the process of safety and security requirements definition and verification? This research direction focuses on common modeling techniques that can incorporate both safety and security failures, and often extends model-based engineering body of knowledge and tools to incorporate security requirements in the design process. \end{inparaenum} In section \ref{sec:related-work}, we survey the main results for each research direction. A more thorough survey is presented in \cite{Kriaa2015b,Lyu2019SafetySystems}.

\textbf{Our Contribution}. The work presented in this paper addresses both research directions with a new approach. First, we integrate both safety and security lifecycles based on a rigorous mathematical formulation that captures their interaction. The formulation enables the designer to assess how a security design decision would impact system safety. This is in contrast to existing research work that does not explicitly model the dynamic safety-security lifecycle interaction. Second, we develop an integrated safety-security design lifecycle and show in details how to apply it to a real world design, distinguishing the work from abstract research on risk assessment that does not carry over to the design stage. Finally, our approach is founded on LOPA,  a practical approach that is extensively used in industry, giving the approach the merit for industrial implementation.

The rest of the paper is organized as follows: Section \ref{sec:safety-security-risk-assessment} introduces the background information required for problem setup, including IEC 61511 safety lifecycle and the LOPA method, cyber dependence between control and safety systems, and cyber security risk assessment. Section \ref{sec:CLOPA} proposes a new LOPA mathematical formulation we call "CLOPA", that incorporates failures due to cyber attacks. Section \ref{sec:co-design} proposes an integrated safety-security lifecycle process. Section \ref{sec:case-study} presents a case study for the design of a safety system for a chemical reactor, comparing classical LOPA approach to the proposed CLOPA formulation. Section \ref{sec:related-work} summarizes the related work on safety-security co-design. The work is concluded in section \ref{sec:conclusion}.

\section{Safety and Security Risk Assessment} \label{sec:safety-security-risk-assessment}
There are two main embedded systems that control and safeguard a given physical system, the control system and the safety system. In the process industry, the control systems is referred to as the Basic Process Control System (BPCS), and the safety system is referred to as the Safety Instrumented System (SIS). In practice, both systems typically have a programmable controller architecture with one or more back planes, processor cards, and a variety of input/output interface cards \cite{Gruhn2006}. For larger systems, BPCS and SIS architectures comprise multiple distributed nodes connected via a communication backbone. Figure \ref{fig:BPCS-SIS-LINK} depicts the two systems and their connectivity over a control network. In the following, we briefly discuss the SIS design lifecycle, BPCS and SIS security lifecycles, and their interaction.

\subsection{IEC 61511 Safety Lifecycle Process} \label{sec:safety-lifecycle}
\begin{figure*}[]
\centering
\includegraphics[trim = 0.5cm 0.5cm 0.5cm 0.5cm, clip, scale=0.65]{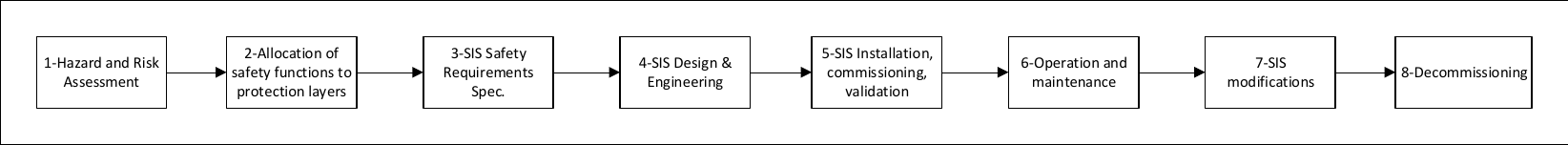}
	\caption{IEC 61511 SIS design lifecycle (adopted from \cite{IEC61511})}
\label{fig:SIS-cycle}
\end{figure*}

Figure \ref{fig:SIS-cycle} shows the Safety Instrumented System (SIS) design lifecycle according to IEC 61511 standard \cite{IEC61511}. The design starts with Hazard \& Risk assessment, where systems hazards are identified. HAzard and OPerability (HAZOP) study, What If analysis, and Fault Tree analysis are the most common methods at this stage \cite{Dunjo2010}. The risk assessment phase ranks each identified risk according to its likelihood and consequence, either quantitatively or qualitatively, and associates a risk ranking for each hazard. The resulting list of hazards and associated risk ranking is used as an input to the second phase focused on the allocation of safety functions to protection layers. This phase deals only with hazards that exceed a threshold risk rank that an organization is willing to accept. For each hazardous scenario, there is a Target Mitigated Event Likelihood (TMEL) measure that is defined based on the risk rank. The purpose of this phase is to check if the TMEL is met with existing protection layers. If not, an additional protection layer is recommended, often in the form of a new Safety Instrumented Function (SIF) with a predefined Safety Integrity Level (SIL) to cover the gap to the TMEL. The safety instrumented function comprises one or more sensors, a logic solver, and one or more actuators. The logic solver is commonly referred to as the Safety Instrumented System (SIS). An example SIF is illustrated in Figure \ref{fig:Reactor P&ID} for an overflow hazardous scenario of a reactor system, which will be discussed in details in section \ref{sec:case-study}. Risk Matrix, Risk Graph, and Layer Of Protection Analysis (LOPA) are the most commonly used methods for the allocation of safety functions to protection layers \cite{Gruhn2006}.

\tikzstyle{block} = [draw, fill= light-gray, rectangle, minimum height=2em, minimum width=22em]
\tikzstyle{clause} = [draw, fill=white, rectangle, rounded corners, minimum height=1em, minimum width=5em]
\tikzstyle{numbox} = [draw, fill=white, rectangle, minimum height=1em, minimum width=1em]

The third phase is the development of the Safety Requirements Specification (SRS), which documents all the functional and timing requirements for each SIF. The fourth phase is the detailed design and engineering. Phases 5 to 8 are concerned with system installation and commissioning, operation, modification, and decommissioning. Phase 2 is where the CPS control and safety systems are considered in the risk assessment process. Therefore, we study this phase in depth in this paper. Since LOPA is the predominant approach for this phase, we limit our discussion to LOPA methodology. Other approaches could be adopted in a similar way.

The underlying assumption in LOPA analysis is that all protection layers, including the new SIF, are independent. In other words, if one layer failed, this does not increase or decrease the likelihood of failure of the other layers. This assumption simplifies the mathematical analysis significantly, as it allows the multiplication of individual probabilities to obtain the required joint probability. The simplicity of LOPA calculations is probably one of the key reasons behind its widespread adoption by industry. Unfortunately, when cyber security is considered as a potential failure in LOPA analysis, the independence assumption between the control and safety systems no longer holds, as explained in the next section.

\subsection{Control and Safety Systems Cyber Dependence} \label{sec:BPCS-SIS-DEPENDENCE}
Each of the control and safety systems has two modes of failure; BPCS physical failures, $B_p$, BPCS security failure, $B_c$, SIS physical failure, $S_p$, and SIS security failure $S_c$. For physical failures, IEC 61511 standard strongly recommends complete separation between the control and safety systems of any plant. This separation includes sensors, computing devices, as well as final elements such as valves and motors. Separation also includes any common utility such as power supplies. The industry adopted this separation principle, hence BPCS and SIS physical failures could be accurately assumed to be independent, i.e., $P[B_p,S_p]=P[B_p]P[S_p]$.

\begin{figure}[]
	\centering
	\includegraphics[scale=0.4,trim=0 0.5cm 0 0.65cm, clip]{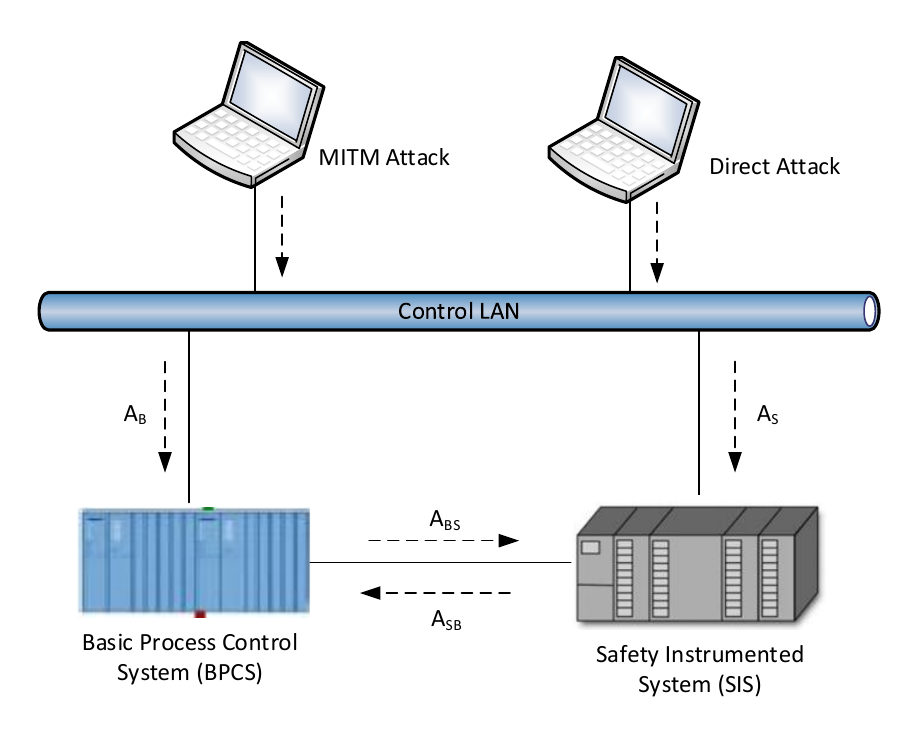}
	\caption{A snapshot of an industrial control system architecture showing BPCS-SIS connectivity and potential attack vectors.}
	\label{fig:BPCS-SIS-LINK}
\end{figure}

One exception to the separation between BPCS and SIS is the cyber communication link between the control and safety systems. Figure \ref{fig:BPCS-SIS-LINK} is a snapshot of a typical industrial control system architecture showing the communication link between the BPCS and SIS. BPCS-SIS communication could be over the control LAN or via a dedicated point to point serial link. The communication protocol is typically an open standard such as Modbus or DNP3 \cite{swales1999open,Fovino2010}. This type of communication exists in many industrial installations to exchange plant data, as the data from field devices connected to the safety system is not accessible from the BPCS and vice versa. Given this architecture, we can define two attack vectors for SIS compromise: \begin{inparaenum}[(1)] \item a direct attack that exploits an existing controller vulnerability could be launched against the SIS node. This could be via any node on the control LAN or using Man In The Middle (MITM) attack that exploits the BPCS-SIS communication. We designate this attack event by $A_S$ in Figure \ref{fig:BPCS-SIS-LINK}, and \item by compromising the BPCS first then exploiting the BPCS-SIS link to compromise the SIS. We designate this pivot attack by the sequence of events $A_B$ and $A_{BS}$ in Figure \ref{fig:BPCS-SIS-LINK}\end{inparaenum}. Further, we designate the attack event from the SIS to the BPCS by $A_{SB}$. The attack sequence $A_B \rightarrow A_{BS}$ may be easier if the SIS is highly secured such that direct attack may be infeasible. This is particularly true if we consider the fact that the BPCS is a trusted node to the SIS.

The above analysis shows a clear dependency between the control and safety systems that violates the original LOPA independence assumption. The security failures for the BPCS and SIS are no longer independent due to the data communication coupling. We can formulate the different security failure probabilities as in (\ref{eq:BPCS-Cyber-Failure-Prob}) to (\ref{eq:BPCS-SIS-Joint-Cyber-Failure-Prob}) using basic probability laws with the aid of Figure \ref{fig:BPCS-SIS-LINK}, where $P[A_i]$ is interpreted as the probability of success of attack $A_i$. Furthermore, the considered attack $A_i$ should have the impact of stopping the BPCS or SIS from performing its intended control or safeguard function as related to the hazard under study. This is important because not all attacks that exploit controller vulnerabilities result in a process hazard. Therefore, the attacks considered represent a subset of the complete set of attacks that could exploit the BPCS or SIS vulnerabilities. Accordingly, from hereafter, $P[A_B], P[A_S], P[A_{BS}], P[A_{SB}]$ refer to the relevant attacks that cause a process hazard. This concept is revisited throughout the paper and is made more clear in the case study when sample attacks are presented. For a case study example on the fact that not all cyber failures have a system reliability consequence, we refer the interested reader to \cite{faza2009reliability} for a study on the impact of different software failure modes on system reliability for the electric power grid domain. Finally, it can be easily shown that if the BPCS-SIS communication link does not exist, or fully secured, then $P[A_{SB}]=P[A_{BS}]=0$, and equation (\ref{eq:BPCS-SIS-Joint-Cyber-Failure-Prob}) reduces to the independent case $P[S_c,B_c]=P[S_c]P[B_c]$.
\begin{align}
P[B_c] &= P[A_B] + P[A_S]P[A_{SB}] - P[A_B]P[A_S]P[A_{SB}]  \label{eq:BPCS-Cyber-Failure-Prob} \\
P[S_c] &= P[A_S] + P[A_B]P[A_{BS}] - P[A_B]P[A_S]P[A_{BS}]  \label{eq:SIS-Cyber-Failure-Prob} \\
P[S_c,B_c] &= P[A_B](P[A_S] + P[A_{BS}]) + P[A_S]P[A_{SB}] \nonumber \\ &- P[A_B]P[A_S](P[A_{BS}]+P[A_{SB}]) \label{eq:BPCS-SIS-Joint-Cyber-Failure-Prob}
\end{align}


\subsection{Cyber Security Risk Assessment}
The calculation of the probability of cyber attacks $A_S,A_B$, and $A_{BS}$ could be performed during the cyber security risk assessment process. This requires a detailed specification of the BPCS and SIS and their connectivity, including the embedded system hardware, operating system, running software services, as well as the network connectivity. According to NIST SP 800-30 standard, "Guide for Conducting Risk Assessments", the cyber security lifecycle process stages are: \begin{inparaenum}[(1)] \item asset identification, where the particular cyber components and their criticality levels are identified, \item vulnerability identification, along with the associated threats and attack vectors, \item development of relevant attack trees for each attack scenario identified, \item penetration testing to validate the vulnerability findings and attack scenarios and to help estimating the effort and probability for individual attack steps for each scenario, and \item risk assessment to identify the scenarios with unacceptable risk \cite{stoneburner2002}. \end{inparaenum} Figure \ref{fig:codesign} shows the BPCS and SIS cyber security lifecycles. In this work, we follow the same cyber security lifecycle, but with the physical process as the main focus. Therefore, for asset identification, the cyber component criticality is primarily identified by its failure impact on the operation of the connected physical component. Similarly, for vulnerability identification, threats and attack vectors are filtered by their impact on the physical process. Attacks that do not disturb the controlled process are ignored as they have no direct impact on the process safety. In addition, such impacts take place with much higher probability at the corporate network level so they can be ignored with minimum impact on the risk assessment at the control network level. For more detailed discussion on process-driven attack identification, we refer the reader to \cite{Tantawy2020}.

For the presented architecture, the BPCS and SIS are the critical components in direct contact with the process. The calculation of the required BPCS and SIS security failure probabilities could be typically carried out with the aid of attack trees \cite{Moore2001}. The attack tree enumerates all possible routes to compromise the system, and each edge is assigned a probability representing the likelihood of the associated event. Using basic probability laws, the overall probability of a system compromise could be calculated. Section \ref{sec:case-study} presents an example of such calculation. We re-emphasize the fact that the scope of cyber security risk assessment and attack trees in this case will be limited to attacks targeting the physical process to cause a process hazard. Although information security attacks with objectives such as stealing information is possible, most of this information is already available at the corporate network level, and an attacker who penetrates down to the control network level to compromise an BPCS or SIS will conceivably have the goal of physical process attack. 

\section{CLOPA: LOPA with Security Failures} \label{sec:CLOPA}
\subsection{Mathematical Formulation}
In risk assessment, an initiating event is an unplanned event that when occurring may lead to a hazard. Examples of initiating events include equipment failure, human error, and cyber attacks. A system hazard will take place if one or more of the initiating events occur and all the associated protection layers against that hazard fail simultaneously. The main objective of LOPA is to calculate the expected number of hazardous events per time interval and compare it to the TMEL. We designate the random variable representing the number of events per unit time for a specific initiating event by $N$, the random variable representing the simultaneous failure of protection layers when the initiating event occurs by $L$, where $L$ is Bernoulli-distributed with success probability $p$, and the random variable representing the number of hazards per time interval by $H$. We then have:
\begin{align}
H = \sum_{i=1}^N \mathbbm{I}_E(l_i)
\end{align}
where $\mathbbm{I}$ is the indicator function, and the set $E=\{l:l_i=1, i=1:N\}$. We note that for a given $N=k$, $H$ is a binomially-distributed random variable with expected value $E[H|N=k] = kp$. Therefore:
\begin{align}
E[H] &= \sum_{k=0}^\infty E[H|N=k]P[N=k] \nonumber \\
     &= p \sum_{k=0}^\infty kP[N=k] = p E[N] = p \lambda
\label{eq:Expected-Hazards}
\end{align}
where $\lambda$ represents the expected value of the number of initiating events per unit time, $N$. Although $N$ is typically modeled by a Poisson random variable in reliability engineering, we do not assume any specific distribution in the analysis. This is particularly important because some initiating events considered in the paper, such as security failures, are not accurately modeled by a Poisson distribution.

Equation (\ref{eq:Expected-Hazards}) is the underlying mathematical concept behind LOPA analysis. Essentially, for each initiating event, the likelihood $\lambda$ is estimated from field data, and the probability of simultaneous failure of all protection layers is specified. Finally, the expected number of hazards per unit time, $E[H]$, considering all initiating events, is estimated and compared to the pre-specified TMEL. If $E[H] > $ TMEL, then a safety instrumented system is required with a probability of failure on demand $P[S_p]$ (or equivalently a Risk Reduction Factor RRF $=1/P[S_p]$) that achieves $E[H] \leq $ TMEL.

In order to express the LOPA formula in (\ref{eq:Expected-Hazards}) in terms of all protection layers, including the BPCS and SIS, we introduce some mathematical notation. We designate the set of initiating events for a given hazardous scenario by $\mathcal{I}=\{I_1, I_2, \ldots, I_n,B_p,\mathcal{A}_r \}$, where $n$ is the number of possible initiating events excluding BPCS failures, $B_p$ denotes BPCS physical failure event, and $\mathcal{A}_r$ denotes the set of attacks relevant to the hazard under study. We express the associated set of event likelihoods by $\Lambda = \{ \lambda_1, \lambda_2,\ldots,\lambda_n, \lambda_p, \lambda_c \}$. Further, we denote the set of all possible protection layers by $\mathcal{L}=\{ L_1, L_2, \ldots, L_m \}$, where $m$ is the number of existing protection layers, excluding the BPCS and SIS. BPCS protection is denoted by $B$, and SIS protection is denoted by $S$. For each initiating event $i$, there is a subset of protection layers $\mathcal{L}_i \subseteq \mathcal{L} \cup \{B,S\}$ that could stop the propagation of a hazard from causing its consequences. Table \ref{tab:LOPA-TABLE-SAMPLE} shows a sample LOPA table using the introduced terminology.

\begin{table}[]
\centering
\begin{tabular}{p{1.5cm} p{1.5cm} p{0.3cm} p{0.3cm} p{0.3cm} p{1cm} p{0.5cm}}
	\hline\hline
	Initiating Event			& Likelihood $\lambda_i$ (/yr)	& $L_1$ & \ldots & $L_m$ & BPCS (B)	& TMEL  \\ \hline\hline
	$I_1$			& $\lambda_1$ 	& \multicolumn{3}{c}{$ \leftarrow P[\mathcal{L}_1]$ $\rightarrow$} & $P[B]$ & $10^{-x}$ \\ 
	\ldots	& \ldots			&  &  & & \ldots &  \\ 
	$I_n$	& $\lambda_n$		& \multicolumn{3}{c}{$ \leftarrow P[\mathcal{L}_n]$ $\rightarrow$}& $P[B]$ &  \\ 
	$B_p$				& $\lambda_p$			& \multicolumn{3}{c}{$ \leftarrow P[\mathcal{L}_B]$ $\rightarrow$} & 1 & \\ 
	$\mathcal{A}_r$				& $\lambda_c$		& \multicolumn{3}{c}{$ \leftarrow P[\mathcal{L}_B]$ $\rightarrow$}& $P[B]$ & \\ \hline
\end{tabular}
\caption{Sample LOPA Table. $P[\mathcal{L}]$ refers to the combined probability of failure of protection layers applicable to the initiating event from $L_1$ to $L_m$.}
\label{tab:LOPA-TABLE-SAMPLE}
\end{table}

\subsection{Semantically-relevant Attack Events Formulation} \label{sec:symantec-attack}
We designate the set of all possible attacks against the BPCS by $\mathcal{A}$. For a given hazard under consideration, the subset of attacks that would lead to this hazard, i.e. the contextually or semantically-relevant attacks, is designated by $\mathcal{A}_r$. To estimate the likelihood of relevant cyber attacks (expected number per unit time), $\lambda_c$, we assume an attacker profile with an average rate of launching attacks per unit time $\lambda$. For every launched attack, the attacker is presented with the complete set of attacks $\mathcal{A}$, and selects only one attack $a$ with probability $P[A=a] = \alpha_a$ that is dependent on the attacker profile, such that:
\begin{align}
\sum_{a \in \mathcal{A}}P[A=a] = \sum_{a \in \mathcal{A}}\alpha_a = 1
\end{align}
The likelihood of cyber attack $a \in \mathcal{A}_r$ is then $\lambda_a = \lambda \alpha_a$. This cyber attack has the potential to cause a system hazard if both BPCS and SIS fail jointly to stop the attack (either physical or cyber failure). We designate this probability by $P_a[S,B]$. The exact approach to include every attack $a \in \mathcal{A}_r$ in the LOPA table is to treat each attack as an individual entry, akin to the last row in Table \ref{tab:LOPA-TABLE-SAMPLE}, with initiating event likelihood $\lambda_a$. However, this approach has two main drawbacks: First, the number of attacks could be large and this would grow the LOPA sheet significantly. Second, the treatment of each attack individually would not allow us to utilize attack modeling techniques such as attack trees that model collectively all possible attack paths for one attack objective. Therefore, we adopt an alternative approach where all relevant attacks $a \in \mathcal{A}_r$ could be represented by one entry in the LOPA table. The following lemma summarizes the approximate solution. The proof is included in Appendix \ref{app:lemma-proof}.
\begin{lemma} \label{lemma:ATTACK-AGGREGATION}
Assume a given hazard scenario $H$, a control system BPCS, an average rate of launching attacks against BPCS $\lambda$, Hazard $H$ semantically-relevant attack set $\mathcal{A}_r$, and probability $\alpha_a$ of selecting attack $a \in \mathcal{A}_r$. Then, the impact of all initiating events $a \in \mathcal{A}_r$ on the LOPA calculation could be approximated by a single initiating event with likelihood $\lambda_c = \lambda \sum_{a \in \mathcal{A}_r} \alpha_a$ and a BPCS failure probability with respect to the combined set of attacks $a \in \mathcal{A}_r$, where each attack probability is weighted by the factor $\gamma_a = \alpha_a / \sum_{a \in \mathcal{A}_r} \alpha_a$.
\end{lemma}
The lemma enables us to use attack trees with leaf nodes weighted by $\gamma_a$ to calculate the BPCS security failure probability in response to the combined set of attacks $\mathcal{A}_r$ with likelihood $\lambda \sum_{a \in \mathcal{A}_r} \alpha_a$. For the special case where the cyber attacker profile results in random selection of the attack $a \in \mathcal{A}$, e.g., an attacker with no knowledge about the system, the likelihood reduces to $\lambda |\mathcal{A}_r|/|\mathcal{A}|$ and the leaf node weights reduce to $\gamma_a = 1/|\mathcal{A}_r|, \forall a$. We use this special case in the case study in Section \ref{sec:case-study}.

\subsection{Cyber LOPA Formulation}
With the introduced notation, the expected number of hazards in (\ref{eq:Expected-Hazards}), which should be less than the TMEL, could be expanded as:
\begin{align}     
E[H] = & P[S,B] \left( \sum_{i=1}^n \left( \lambda_i P[\mathcal{L}_i] \right) + \lambda_c P[\mathcal{L}_B] \right) + \nonumber \\ 
& \lambda_p P[S] P[\mathcal{L}_B] \leq \text{TMEL}
\label{eq:LOPA-GENERAL}
\end{align}
where $\mathcal{L}_B$ is the set of protection layers for BPCS physical or security failure event, and we assume that all protection layers are independent from the BPCS and SIS, while keeping the dependence between the BPCS and SIS. In addition, higher order probability terms resulting from multiple initiating events are ignored due to their insignificance.

To calculate the joint failure probability $P[S,B]$, we use basic probability laws and the fact that the BPCS and SIS have both physical and cyber modes of failure as explained in Section \ref{sec:BPCS-SIS-DEPENDENCE} to obtain:
\begin{align}     
P[S,B] &= P[S_p] \left( P[B_p](1-P[B_c]-P[S_c])+P[B_c] \right) \nonumber \\
&+ P[S_c,B_c] \left( 1 - P[S_p] - P[B_p] + P[S_p]P[B_p] \right) \nonumber \\
&+ P[S_c]P[B_p] \label{eq:Joint-Prob}
\end{align}


Substituting (\ref{eq:Joint-Prob}) in (\ref{eq:LOPA-GENERAL}), we obtain the general LOPA equation in (\ref{eq:LOPA-General-Expanded}). We call this expanded version of LOPA hereafter \emph{\textbf{CLOPA}}, standing for Cyber LOPA.
\begin{align}     
P[S_p] \leq \frac{\beta - (\alpha_1 P[S_c] + \alpha_2 P[S_c,B_c])}{\alpha_1 - \alpha_1 P[S_c] + \alpha_2 P[B_c] - \alpha_2 P[S_c,B_c]}
\label{eq:LOPA-General-Expanded}
\end{align}
where:
\begin{align}
\alpha_1 &= P[B_p] \left( \sum_{i=1}^n \left( \lambda_i P[\mathcal{L}_i] \right) + \lambda_c P[\mathcal{L}_B] \right) + \lambda_p P[\mathcal{L}_B] \label{eq:alpha1}\\
\alpha_2 &= (1-P[B_p]) \left( \sum_{i=1}^n \left( \lambda_i P[\mathcal{L}_i] \right) +  \lambda_c P[\mathcal{L}_B] \right) \label{eq:alpha2}\\
\beta &= \text{TMEL} \label{eq:beta}
\end{align}

In order to define the CLOPA formula in terms of the actual design variables $P[A_S]$ and $P[A_{BS}]$ that represent the probability of security failures of actual CPS components, we substitute  (\ref{eq:BPCS-Cyber-Failure-Prob}) - (\ref{eq:BPCS-SIS-Joint-Cyber-Failure-Prob}) into (\ref{eq:LOPA-General-Expanded}) to obtain:
\begin{align}     
P[S_p] \leq \frac{\beta - \gamma_1 P[A_S] - \gamma_2 P[A_{BS}](1 - P[A_S])}{\gamma_3 - \gamma_3 P[A_S] - \gamma_2 P[A_{BS}] ( 1 - P[A_S])}
\label{eq:LOPA-Symbolic-CPS-COMPONENTS}
\end{align}
where:
\begin{align}
\gamma_1 &= \alpha_1 + \alpha_2 [P[A_B] + P[A_{SB}] (1 - P[A_B])] \label{eq:gamma1}\\
\gamma_2 &= (\alpha_1 + \alpha_2) P[A_B] \label{eq:gamma2}\\
\gamma_3 &= \alpha_1 + \alpha_2 P[A_B] \label{eq:gamma3}
\end{align}
Equation (\ref{eq:LOPA-Symbolic-CPS-COMPONENTS}), along with (\ref{eq:alpha1}) - (\ref{eq:beta}) and (\ref{eq:gamma1}) - (\ref{eq:gamma3}), represent the general CLOPA formulation to design the safety instrumented system. It represents an upper bound on the probability of physical failure for the safety system in terms of the security failure probabilities, showing clearly the coupling between the safety system and security system design. The design of the safety instrumented system should satisfy (\ref{eq:LOPA-Symbolic-CPS-COMPONENTS}), where the design variables are $P[S_p],P[A_S]$, and $P[A_{BS}]$. The rest are model parameters that are predetermined, including the BPCS failure marginal probabilities. This is because the BPCS design is independent of the SIS design, and usually takes place earlier in the engineering design cycle. Note that we assume here that $P[A_{SB}]$ is a known parameter. This is because by completely defining the BPCS and its hardware and software specifications, the probability of a cyber attack compromising process safety could be estimated, even though the SIS is not yet completely defined. Table \ref{tab:Model-Parameters} in the appendix summarizes the model variables, parameters, and how the model parameters are calculated.

It should be noted that with existing LOPA methodology, security failures are ignored, i.e., $P[A_B]=P[A_S]=P[A_{BS}]=P[A_{SB}]=0$. Substituting these zero values in (\ref{eq:LOPA-Symbolic-CPS-COMPONENTS}), we obtain the classical LOPA formulation:
\begin{align}     
P[S_p] \leq \frac{\beta}{\alpha_1} = \frac{\text{TMEL}}{ P[B_p] \sum_{i=1}^n \left( \lambda_i P[\mathcal{L}_i] \right) + P[\mathcal{L}_B] (\lambda_p + \lambda_c)}
\label{eq:LOPA-SPECIAL}
\end{align}

\subsection{Design Space}
\begin{figure}[]
\centering
\includegraphics[scale=0.3,trim={0 0cm 0 0.4cm},clip]{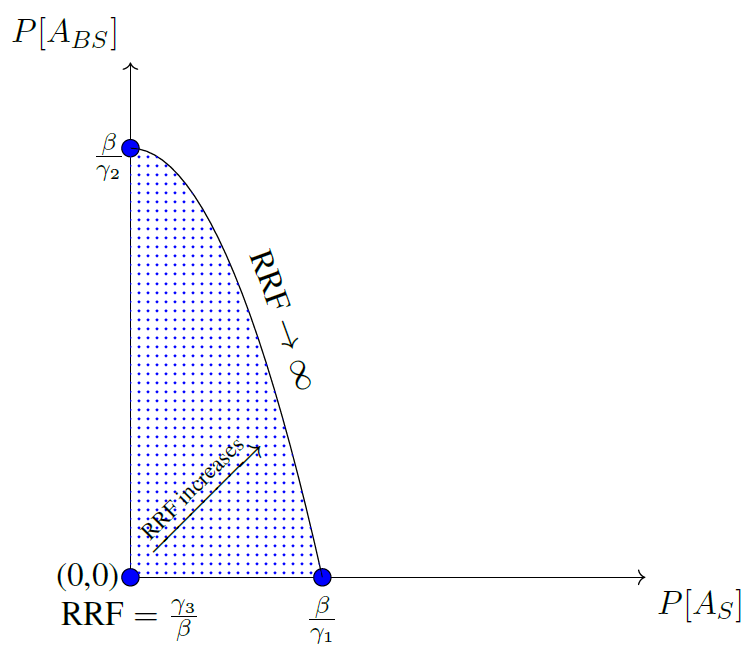}
\caption{CLOPA Design Region (shaded). Any point in the shaded region results in a feasible SIS. Points near the boundary requires a SIS with a very high RRF value, hence difficult to obtain in practice.}
\label{fig:design-region-cyber-prob}
\end{figure}

Using the fact that $P[S_p] \geq 0$ for a realizable safety system in (\ref{eq:LOPA-Symbolic-CPS-COMPONENTS}), we obtain:
\begin{align}
\gamma_1 P[A_S] + \gamma_2 P[A_{BS}] - \gamma_2 P[A_S]P[A_{BS}] \leq \beta        \label{eq:design-constraint}
\end{align}
Figure \ref{fig:design-region-cyber-prob} shows the shaded region defined by the inequality in (\ref{eq:design-constraint}). The boundary curve is defined by (\ref{eq:design-constraint}) when equality holds:
\begin{align}
P[A_{BS}] = \frac{\beta}{\gamma_2} \left( \frac{1-\left(\frac{\gamma_1}{\beta} \right) P[A_S]}{1-P[A_S]} \right) \label{eq:design-boundary}
\end{align}
The first order derivative of the boundary curve is negative for $\gamma_1/\beta > 1$ and positive otherwise. Since $\gamma_1 > \beta$ to require a safety system (proof is straightforward by inspecting equations (\ref{eq:alpha1}),(\ref{eq:alpha2}), (\ref{eq:gamma1}), and (\ref{eq:LOPA-SPECIAL})), the boundary curve is concave as in Figure (\ref{fig:design-region-cyber-prob}b). We note that any point in the shaded region results in a feasible SIS. Points on the boundary curve result in $P[S_p] = 0$, or equivalently $\text{RRF} \rightarrow \infty$. Points closer to the boundary would have high values for the $\text{RRF}$, requiring a very highly reliable SIS that may not be achievable in practice. Points closer to the origin result in lower RRF. It can be easily shown that the contour lines for (\ref{eq:LOPA-Symbolic-CPS-COMPONENTS}), where $P[S_p] = C$, could be expressed as:
\begin{align}
P[A_{BS}] = \frac{C \gamma_3 - \beta}{\gamma_2 (C-1)} \left( \frac{1 - \left(\frac{C \gamma_3 - \gamma_1}{C \gamma_3 - \beta}\right) P[A_S]}{1-P[A_S]} \right)
\label{eq:Contour-Lines}
\end{align}
The contour line that represents the design boundary in Figure \ref{fig:design-region-cyber-prob} can be derived from (\ref{eq:Contour-Lines}) by setting $C = 0$.

We can extract several information from this graph: \begin{inparaenum} [(1)]\item the maximum probability of security failure for the safety system by directed attacks is $\beta/\gamma_1$. This probability results in an unrealizable safety system, as the required RRF $\rightarrow \infty$. \item The maximum probability of security failure for the safety system by pivot attack via the BPCS is $\beta/\gamma_2$. Likewise, this probability does not result in a realizable safety system. Finally \item The minimum value of RRF is achieved at the origin for a perfectly secured safety system where $P[A_S] = P[A_{BS}] = 0$, with RRF given by \end{inparaenum}:
\begin{align}
P[S_p]_{\max} = \frac{\beta}{\gamma_3}, \quad \text{RRF}_{\min}= \frac{\gamma_3}{\beta}
\end{align}
Clearly, points outside the shaded region result in non-realizable SIS. This result re-emphasizes the interplay between the safety and security systems of a cyber physical system.

The design space highlights the major difference between LOPA and CLOPA. In LOPA, the SIS requirement is related to reliability in the form of the required safety integrity level. In CLOPA, an additional requirement for the SIS is its security resilience, in the form of an upper bound on the probability of a security failure (cyber attack success), either directly or indirectly via the BPCS.

\subsection{Classical LOPA Error}
To obtain the error resulting from using classical LOPA, we subtract (\ref{eq:LOPA-SPECIAL}) from (\ref{eq:LOPA-Symbolic-CPS-COMPONENTS}) to obtain:
\begin{align}
e_{\text{RRF}} = \frac{\zeta_1 + \zeta_2 P[A_S] + \zeta_3 P[A_{BS}] (1 - P[A_S])}{\beta \left[ \beta - \gamma_1 P[A_S] - \gamma_2 P[A_{BS}] (1- P[A_S]) \right]}
\label{eq:LOPA-Error}
\end{align}
where:
\begin{align}
\zeta_1 &= \beta (\gamma_3 - \alpha_1) = \beta \alpha_2 P[A_B] \label{eq:zeta1}\\
\zeta_2 &= \alpha_1 \gamma_1 - \beta \gamma_3 \label{eq:zeta2}\\
\zeta_3 &= \gamma_2 (\alpha_1 - \beta) \label{eq:zeta3}
\end{align}
The minimum error occurs for a perfectly secured safety system, i.e., $P[A_S]=P[A_{BS}]=0$:
\begin{align}
\min e_{\text{RRF}} &= P[A_B] \left(\frac{\alpha_2}{\beta} \right) 
\end{align}
The error will be zero, i.e., classical LOPA result matches CLOPA, if the probability of BPCS security failure via a direct attack is zero.

\section{Safety-Security Co-Design} \label{sec:co-design}
\subsection{Design Process}
The current industrial practice is to perform safety and security risk assessments independently, treating the physical and cyber components of a CPS as two separate entities. As illustrated in section \ref{sec:CLOPA}, accurate safety risk assessment requires knowledge about the cyber components and their security failure probabilities. Formally, the objective is to design a safety instrumented system architecture $\mathcal{A}$ that satisfies (\ref{eq:LOPA-Symbolic-CPS-COMPONENTS}) in terms of both physical and security failure probabilities. Suppose that the architecture $\mathcal{A}$ could be represented by a set of design variables represented by the vector $\mathbf{x}$. If we can relate the physical and security failure probabilities to the vector $\mathbf{x}$ by $P[A_S] = f(x), P[A_{BS}] = g(x), P[S_p] = h(x)$, then we can use these functions to substitute the relevant probabilities in (\ref{eq:LOPA-Symbolic-CPS-COMPONENTS}) and our design problem will be to find a set of values for the vector $\mathbf{x}$ that satisfies the CLOPA constraint (\ref{eq:LOPA-Symbolic-CPS-COMPONENTS}). Unfortunately, this design approach is not followed by industry for several reasons. First, abstracting a given architectural design $\mathcal{A}$ into a set of design variables is a very difficult task, not to mention that these design variables have to be linked to both physical and security failures. Second, finding an exact or approximate representation of the functions $f(.), g(.)$, and $h(.)$ that relate the failure probabilities to the design variables may not be possible, as it is not always clear how a design decision would result in a higher or lower probability of failure. Finally, even if we were able to make a perfect modeling, the resulting problem to solve may turn into a discrete optimization problem that is not possible to solve in polynomial time.

Due to these modeling limitations, the current industrial practice to design safety instrumented systems (excluding cyber attacks) is to follow an iterative process and rely on engineering judgement during the design process. More precisely, the required risk reduction factor RRF$_d$ is initially calculated, then the engineering design proceeds to achieve RRF$_d$ using both experience and industrial standard guidelines \cite{IEC61511}. After the design is completed, design verification is conducted to calculate the risk reduction factor of the proposed design RRF$_v$. If The resulting RRF$_v \geq$ RRF$_d$, then the design stops. Otherwise, the design is refined until the condition RRF$_v \geq$ RRF$_d$ is satisfied. In the following, we will adopt the same iterative design approach for CLOPA.

\begin{figure}[]
\centering
\includegraphics[trim = 0.7cm 0.8cm 0.65cm 0.65cm, clip, scale=0.6]{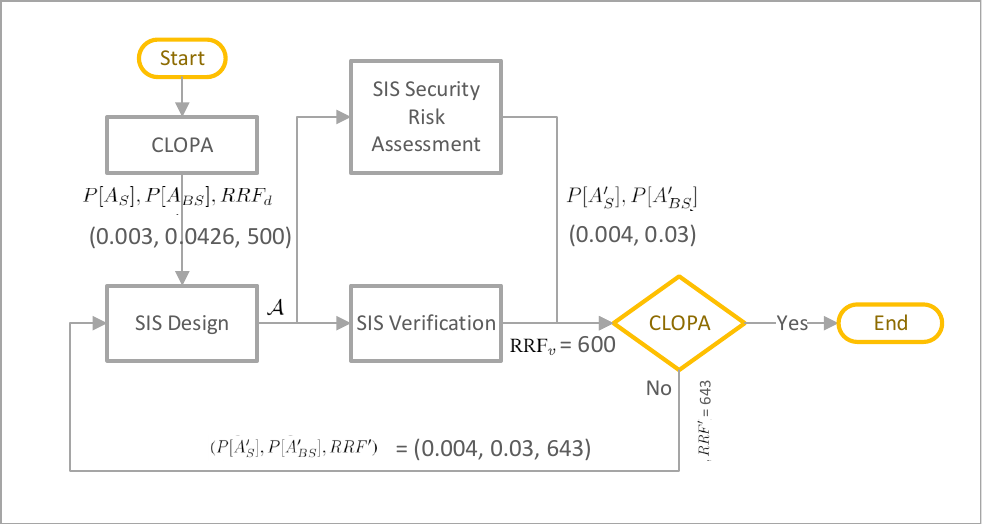}
\caption{CLOPA Iterative Design Process - CSTR case study design values are shown}
\label{fig:Feedback-Design}
\end{figure}

Figure \ref{fig:Feedback-Design} illustrates the iterative design process. We start with initial values ($P[A_S], P[A_{BS}]$, RRF$_d$) that satisfy CLOPA constraint in (\ref{eq:LOPA-Symbolic-CPS-COMPONENTS}). We then proceed with SIS design to produce an architecture $\mathcal{A}$. The architecture is then verified to estimate its probability of failure on demand, or equivalently its risk reduction factor RRF$_v$. The architecture is also used to carry out a security risk assessment to estimate the probability of security failures $P[A^\prime_S]$ and $P[A^\prime_{BS}]$. If the new set of obtained values ($P[A^\prime_S], P[A^\prime_{BS}], $RRF$_v$) still satisfy the CLOPA equation, the design stops. Otherwise, a new iteration will start to adjust the design in order to achieve the CLOPA constraint. This adjustment could be by adding more security controls, or by increasing the reliability of the system using fault tolerant techniques. Algorithm \ref{alg:DesignProcess} summarizes the iterative design process.

One question is how can we choose the initial values for RRF $, P[A_S], P[A_{BS}]$? This initial design point could be selected with the aid of the design contour plot as in figure \ref{fig:design-region-cyber-prob}, where the design point strikes a balance between security and reliability. What is \emph{reasonable} regarding the risk reduction factor is well-documented in the standards using Safety Integrity Levels (SIL), and the extra cost to move from one SIL to a higher SIL is well quantified in industry. What is not very clear, though, is what is an achievable value for the probability of security failures. This is still not a well-developed field, and the argument of how to assess such probabilities is still going on in the research community.

Another important question is whether there is any formal guarantees that Algorithm \ref{alg:DesignProcess} will terminate. To answer this question, we need to know, or at least approximate, how the architectural design $\mathcal{A}$ impacts the RRF, $P[A_S]$, and $P[A_{BS}]$. As pointed out earlier, this is very hard in practice. Without such relationship, the question of convergence to a solution for algorithm termination cannot be precisely answered. However, in practice, modifying the SIS design to increase the RRF is usually done by changing sensor and actuator configuration or reliability figures, as they are often the weakest links in the reliability chain, while the logic solver is minimally changed \cite{IEC61511}. Accordingly, for all practical purposes, we can assume that the design process will converge after few runs.

\begin{algorithm}[]
\SetKwData{Left}{left}\SetKwData{This}{this}\SetKwData{Up}{up}
\SetKwFunction{Union}{Union}\SetKwFunction{FindCompress}{FindCompress}
\SetKwInOut{Input}{input}\SetKwInOut{Output}{output}
\SetKwRepeat{Do}{do}{while}
\Input{BPCS}
\Output{$\mathcal{A}$, $\bm{\theta}_S$}
$([P[A_B], P[A_{SB}]) \leftarrow$ \texttt{BPCS-SecCycle}(BPCS) \;
$\bm{\theta}_B \leftarrow (P[A_B], P[A_{SB}])$ \;
$\left(P[A_S], P[A_{BS}], \text{RRF}_d \right) \leftarrow$ \texttt{DesignContour}($\bm{\theta}_B$) \;
$\bm{\theta}_S \leftarrow \left(P[A_S], P[A_{BS}], \text{RRF}_d \right)$ \;
\Do{$\text{RRF}_v < \text{RRF}^\prime$}
{
    $\mathcal{A}$ $\leftarrow$ \texttt{SIS-SafeCycle}($\bm{\theta}_S$) \;
    $\text{RRF}_v \leftarrow$ \texttt{SIS-Verify($\mathcal{A}$)} \;
    $(P[A^\prime_S], P[A^\prime_{BS}]) \leftarrow$ \texttt{SIS-SecCycle}($\mathcal{A}$) \;
    $\text{RRF}^\prime \leftarrow$ \texttt{CLOPA}$\left(P[A^\prime_S], P[A^\prime_{BS}] , \bm{\theta}_B \right)$ \;
    $\bm{\theta}_S \leftarrow \left(P[A^\prime_S], P[A^\prime_{BS}], \text{RRF}^\prime \right)$ \;
}
\algorithmicreturn{ $\mathcal{A}$, $\bm{\theta}_S$} \;
\caption{Integrated Safety-Security Lifecycle Design Algorithm}
\label{alg:DesignProcess}
\end{algorithm}

\subsection{Integrated Safety-Security Lifecycle} \label{sec:integrated-lifecycle}
As the analysis in this paper shows a clear coupling between safety and security design requirements, we propose the integrated lifecycle in Figure \ref{fig:codesign}. In the following, We present a brief description of the lifecycle steps in the order of their execution, according to the numbering labels in Figure \ref{fig:codesign}. 
\tikzstyle{block} = [draw, fill=white, rectangle, rounded corners, minimum height=2em, minimum width=5em]
\begin{enumerate}[label=\protect\circled{\arabic*},wide, labelindent=0pt, itemsep=0.5ex]
\item SIS Safety Lifecycle - (HAZOP): The first step is to carry the hazard analysis for the physical system, often using HAZOP. This process identifies important assets that may be subject to, or contribute to, risk scenarios. Then, the process identifies all feasible hazards and associated risk ranking, as well as the associated cyber components for each identified hazard. This constitutes an input to the BPCS security lifecycle. If we designate the set of hazards by $\mathcal{H}$, and the set of cyber components by $\mathcal{C}$, then the output from this process is the function $f: \mathcal{H} \mapsto \mathbb{R}$ representing the risk ranking, and the relation $R \subseteq \mathcal{H} \times \mathcal{C}$ representing the cyber components for each hazard.
\item BPCS Cyber Security Lifecycle: The BPCS cyber security lifecycle, including vulnerability analysis, attack tree generation, penetration testing, and risk assessment, is performed on the BPCS. Ideally, the security risk assessment should be carried out for each process hazard scenario identified during HAZOP to identify the relevant vulnerabilities that may cause a process disruption. However, for the given centralized architecture, the BPCS is typically controlling a large number of control loops, hence it may not be necessary to repeat the security risk assessment process for each control loop, as vulnerabilities may be applicable to several hazardous scenarios. The output of this process is the BPCS security failure probabilities $P[A_B]$ and $P[A_{SB}]$.
\item SIS Safety Lifecycle - CLOPA and SIS Design: The first iteration of CLOPA and SIS design will proceed according to Algorithm \ref{alg:DesignProcess} and Figure \ref{fig:Feedback-Design}. The CLOPA calculates the design requirement for the SIS in terms of its reliability as defined by the RRF, and its cyber security resilience as defined by $P[A_S]$ and $P[A_{BS}]$. The SIS design then proceeds according to IEC 61511 standard \cite{IEC61511} to produce an architecture $\mathcal{A}$. The design includes the hardware architecture, redundancy scheme, and software architecture. The specific design architecture can vary across industries and organizations, but the design has to achieve the required RRF, $P[A_S]$ and $P[A_{BS}]$, as calculated by CLOPA. After the design is completed, SIS verification is carried out to calculate the risk reduction factor $\text{RRF}_v$. It should be highlighted that the SIS is one component only of the Safety Instrumented Function (SIF). The SIF includes the sensor, SIS, and the actuator. Therefore, the verification is carried out on the whole SIF. For a detailed discussion on SIS design and verification, the reader is referred to \cite{Gruhn2006}.
\item SIS Cyber Security Lifecycle: Using the resulting SIS design hardware and software architecture $\mathcal{A}$, the SIS security lifecycle is carried out. Since the SIS is not yet implemented at this stage, SIS penetration testing is not possible and hence omitted from the security lifecycle. The output from this process is the SIS security failure probabilities $P[A^\prime_S],P[A^\prime_{BS}]$, derived from SIS vulnerabilities that may lead to a process hazard. It is noted that the SIS security lifecycle at the right of figure \ref{fig:codesign}  proceeds from bottom to top for a better presentation.
\item Safety Lifecycle - CLOPA: The CLOPA calculation is carried out using the values obtained from the safety verification and SIS security lifecycle, $\left( P[A^\prime_S],P[A^\prime_{BS}], \text{RRF}_v \right)$, to verify that the architecture $\mathcal{A}$ satisfies the CLOPA constraint. The process SIS safety lifecycle $\rightarrow$ SIS Cyber security lifecycle $\rightarrow$ CLOPA (designated by the blue arrowed arc in figure (\ref{fig:codesign})) repeats until the CLOPA constraint is satisfied.
\item Installation: The finalized design then moves to the installation phase.
\end{enumerate}

\begin{figure}[]
\centering
\scalebox{0.75}{
\begin{tikzpicture}[node distance=1.5cm and 5cm, on grid, >=latex', every node/.style={scale=0.8}, scale=1]
	\node [block] at (0,0) (AID) {Asset ID};
	\node [block, below of = AID] (VID) {Vuln. ID};
	\node [block, below of = VID] (ATT) {Attack Trees};
	\node [block, below of = ATT] (PEN) {Pen.Testing};
	\node [block,below of = PEN] (RA) {Risk Assess.};
	\draw[->] (AID) -- (VID);
	\draw[->] (VID) -- (ATT);
	\draw[->] (ATT) -- (PEN);
	\draw[->] (PEN) -- (RA);	
	\draw [ thick, draw=black, fill=gray, opacity=0.1]		(-1,0.5) -- (1,0.5) -- (1,-5.25) -- (-1,-5.25) -- (-1,0.5);
	\node at (0,-5.5) {BPCS Security Lifecycle};
	\node[circle,draw=black, fill=white, inner sep=1pt,minimum size=5pt] (st2) at (-1,0.5) {2};
	\node [block, right = 3cm of AID] (HAZ) {HAZOP};
	\node [block, below of = HAZ] (CLOPA) {CLOPA};
	\node [block, below of = CLOPA] (SRS) {SRS};
	\node [block, below of = SRS] (Design) {Design};
	\node [block, below of = Design] (inst) {Installation};
	\draw[->] (HAZ) -- (CLOPA);
	\draw[->] (CLOPA) -- (SRS);
	\draw[->] (SRS) -- (Design);
	\draw[->] (Design) -- (inst);
	\draw [ thick, draw=black, fill=gray, opacity=0.1]		(2,0.5) -- (4,0.5) -- (4,-5.25) -- (2,-5.25) -- (2,0.5);
	\node at (3,-5.5) {SIS Safety Lifecycle};
	\draw [ thick, draw=black, fill=gray, opacity=0.2]		(2.2,-0.75) -- (3.8,-0.75) -- (3.8,-4) -- (2.2,-4) -- (2.2,-0.75);
	\node[circle,draw=black, fill=black, inner sep=0pt,minimum size=5pt] (ST) at (3,1) {};
	\node[circle,draw=black, fill=white, inner sep=1pt,minimum size=5pt] (st1) at (2.3,0.3) {1};
	\node[circle,draw=black, fill=white, inner sep=1pt,minimum size=5pt] (st3) at (2.3,-0.75) {3};
	\node[circle,draw=black, fill=white, inner sep=1pt,minimum size=5pt] (st3) at (2.7,-0.75) {5};	
	\node[circle,draw=black, fill=white, inner sep=1pt,minimum size=5pt] (st3) at (2.3,-4.5) {6};		
	\draw[->] (ST) -- (HAZ);
	\node [block, right = 3cm of HAZ] (RAS) {Risk Assess.};
	\node [block, below of = RAS] (Pens) {Pen. Testing};
	\node [block, below of = Pens] (ATTS) {Attack Trees};
	\node [block, below of = ATTS] (VULS) {Vuln. ID};
	\node [block, below of = VULS] (AIDS) {Asset ID};
	\draw[->] (AIDS) -- (VULS);
	\draw[->] (VULS) -- (ATTS);
	\draw[->] (ATTS) -- (Pens);
	\draw[->] (Pens) -- (RAS);
	\draw [ thick, draw=black, fill=gray, opacity=0.1]		(5,0.5) -- (7,0.5) -- (7,-5.25) -- (5,-5.25) -- (5,0.5);
	\node at (6,-5.5) {SIS Security Lifecycle};
	\node[circle,draw=black, fill=white, inner sep=1pt,minimum size=5pt] (st2) at (5,0.5) {4};		
	\draw[->] (HAZ) -- node [text width=1cm,midway,above ] {($\mathcal{C,H}$)}  (AID);
	\draw[->] (RA) --  node [sloped, anchor=center, text width=1.5cm,above ] {\footnotesize$P[A_B],P[A_{SB}]$}  (CLOPA);
	\draw[->] (Design) --  node [sloped, anchor=center, text width=0.5cm,above ] {\footnotesize $\mathcal{A}$} (AIDS);
	\draw[->] (RAS) -- node [sloped, anchor=center, text width=2.5cm,above ] {\footnotesize $P[A^\prime_B],P[A^\prime_{SB}]$} (CLOPA);
	\draw[->,blue] (4.9,-2.5) arc (0:270:4mm);
	\end{tikzpicture}}
\caption{Integrated Safety and Security lifecycles. The process starts at (1) HAZOP, followed by (2) BPCS complete security lifecycle, then (3) SIS safety lifecycle up to the end of the design stage, followed by (4) complete SIS security lifecycle, (5), CLOPA check, and possibly several iterations of steps (3) (4), and (5), then terminates at the SIS installation stage.}
\label{fig:codesign}
\end{figure}

\section{Integrated Design Example}  \label{sec:case-study}
In this section, we present an integrated design example for a process control system to illustrate the proposed CLOPA and integrated lifecycle. The system described in this section is a real testbed located in Qatar University, and comprises the process simulator and the full plant control system. As the integrated design lifecycle is substantial, with some steps outside the scope of this work (e.g., SIS architectural design and security risk assessment), it is not possible to present the design process in full details. However, we try to focus on the big picture as related to the proposed CLOPA, while discussing briefly each design step. Wherever needed, we refer the reader to relevant references for further details.

\subsection{CPS Description}
\begin{figure}[]
\centering
\includegraphics[scale = 0.4, trim = {0.8cm 0.5cm 2.5cm 0.5cm}, clip]{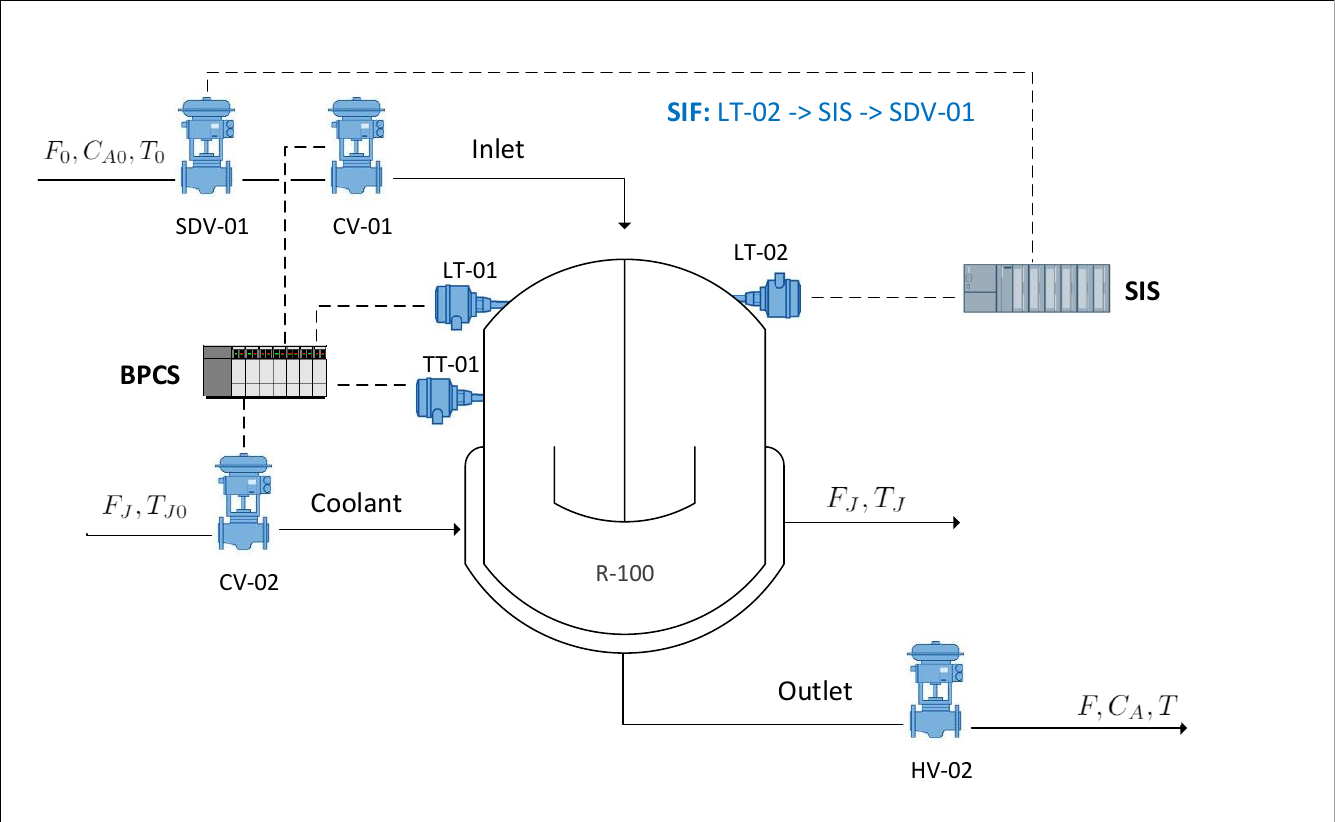}
\caption{Reactor Piping and Instrumentation Diagram (P\&ID). ISA standard symbols are not strictly followed for illustration purposes.}
\label{fig:Reactor P&ID}
\end{figure}

We consider the Continuous Stirred Tank Reactor (CSTR) process illustrated in Figure \ref{fig:Reactor P&ID}. The reactor vessel has an inlet stream carrying the reactant A, an outlet stream carrying the product B, and a cooling stream carrying the cooling fluid into the surrounding jacket to absorb the heat of the exothermic reaction. A first order reaction takes place where a mole fraction of reactant $A$ is consumed to produce product $B$. The process has a level control loop (LT-01 $\rightarrow$ BPCS $\rightarrow$ CV-01) to maintain the liquid level in the reactor, and a temperature control loop (TT-01 $\rightarrow$ BPCS $\rightarrow$ CV-02) to control the reaction rate. The SIF (LT-02 $\rightarrow$ SIS $\rightarrow$ SDV-01) protects the reactor from the overflow hazard, and will be explained later in this section. For more detailed explanation about the process including the state space model, the reader is referred to \cite{Tantawy2019CICN}.

The CSTR process is controlled by the industrial control system shown in Figure \ref{fig:CPS-ARCHI-NIST}, which follows NIST 800-82 standard with one firewall and a DeMilitariZed (DMZ) zone \cite{stouffer2011}. The corporate network cannot communicate directly with the control network. The only allowable information flow paths via the firewall are from the control network to the data logging servers in the DMZ zone, and from the corporate network to the DMZ for information retrieval. The BPCS and SIS have Modbus/TCP communication over the control network \cite{swales1999open}.

\subsection{Integrated Lifecycle}
In the following discussion, we follow the integrated lifecycle in Figure \ref{fig:codesign}, and as per the itemized steps in section \ref{sec:integrated-lifecycle}

\begin{enumerate}[label=\protect\circled{\arabic*},wide, labelindent=0pt, itemsep = 0.5ex]
\item SIS Safety Lifecycle - HAZOP: Table \ref{tab:HAZOP} shows the HAZOP sheet for the CSTR process. Each row contains: \begin{inparaenum}[(1)] \item the possible hazard, \item all possible initiating events for each hazard whether mechanical or electronic failures, \item consequences if the hazard occurred, including safety, financial, and environmental losses, \item existing safeguards that could prevent the hazard from propagating and causing the consequences, and \item the risk rank, which is typically a function of the consequences. \end{inparaenum} There are two identified hazards for the reactor process; high level causing an overflow hazard, and high temperature that may lead to reactor runaway and possible meltdown. Both hazards have high and very high risk rankings, therefore, the two risk scenarios qualify for further LOPA assessment. In the following, we limit our discussion to the high-level hazard scenario only. High temperature hazard could be treated similarly.
\begin{table*}[]
\begin{center}
	\begin{tabular}{ p{1in} p{1.75in} p{2in} p{0.75in} p{0.75in}} 
		\hline\hline
		\textbf{Hazard} & \textbf{Initiating Event} (Cause) & \textbf{Consequences} & \textbf{Safeguards} (IPL) & Risk Rank \\ 
		\hline\hline
		High Level (Reactor overflow) & BPCS failure OR Human error (misaligned valves) & 2 or more fatalities (safety), Product loss (financial), Environmental contamination (environment) & Reactor dike (Mitigation) & High \\
		High Temperature (Reactor Meltdown/explosion) &  BPCS failure OR Coolant inlet control valve fully (partially) closed OR Inlet valve stuck fully open & 10 or more fatalities (safety), Product loss (financial), Environmental contamination (environment) & None & V. High \\
		\hline
	\end{tabular}
\end{center}
\caption{Partial HAZOP sheet for the reactor process}
\label{tab:HAZOP}
\end{table*}
\item BPCS Cyber Security Lifecycle: We need to calculate $P[A_B]$ and $P[A_{SB}]$ for the BPCS,  the probability that the BPCS fails due to a direct attack and a SIS-pivot attack, respectively, in a way that generates the high level process hazard. We conducted vulnerability identification on the CPS network in Figure \ref{fig:CPS-ARCHI-NIST}, constructed the attack trees, and carried out penetration testing to verify the vulnerability findings. We assumed an attacker profile where attacks are selected randomly. The total number of semantically-relevant attacks is found to be $|\mathcal{A}_r|=42$. We assume that relevant attacks represent $10^{-4}$ of all possible attacks, i.e., $|\mathcal{A}_r|/|\mathcal{A}|=10^{-4}$. Therefore, according to Lemma \ref{lemma:ATTACK-AGGREGATION}, we obtain an initiating event likelihood $10^{-4}\lambda$ and a weight factor $\gamma_a = 0.024$ for each attack $a$ at the leaf nodes of the attack tree. As the full details of vulnerability analysis, attack design, and penetration testing are beyond the scope of this paper, we refer the interested reader to \cite{Tantawy2020,Tantawy2021automated}.

\begin{figure}[]
\centering
\includegraphics[trim = 0.5cm 0.5cm 0.5cm 1cm, clip, scale=0.4]{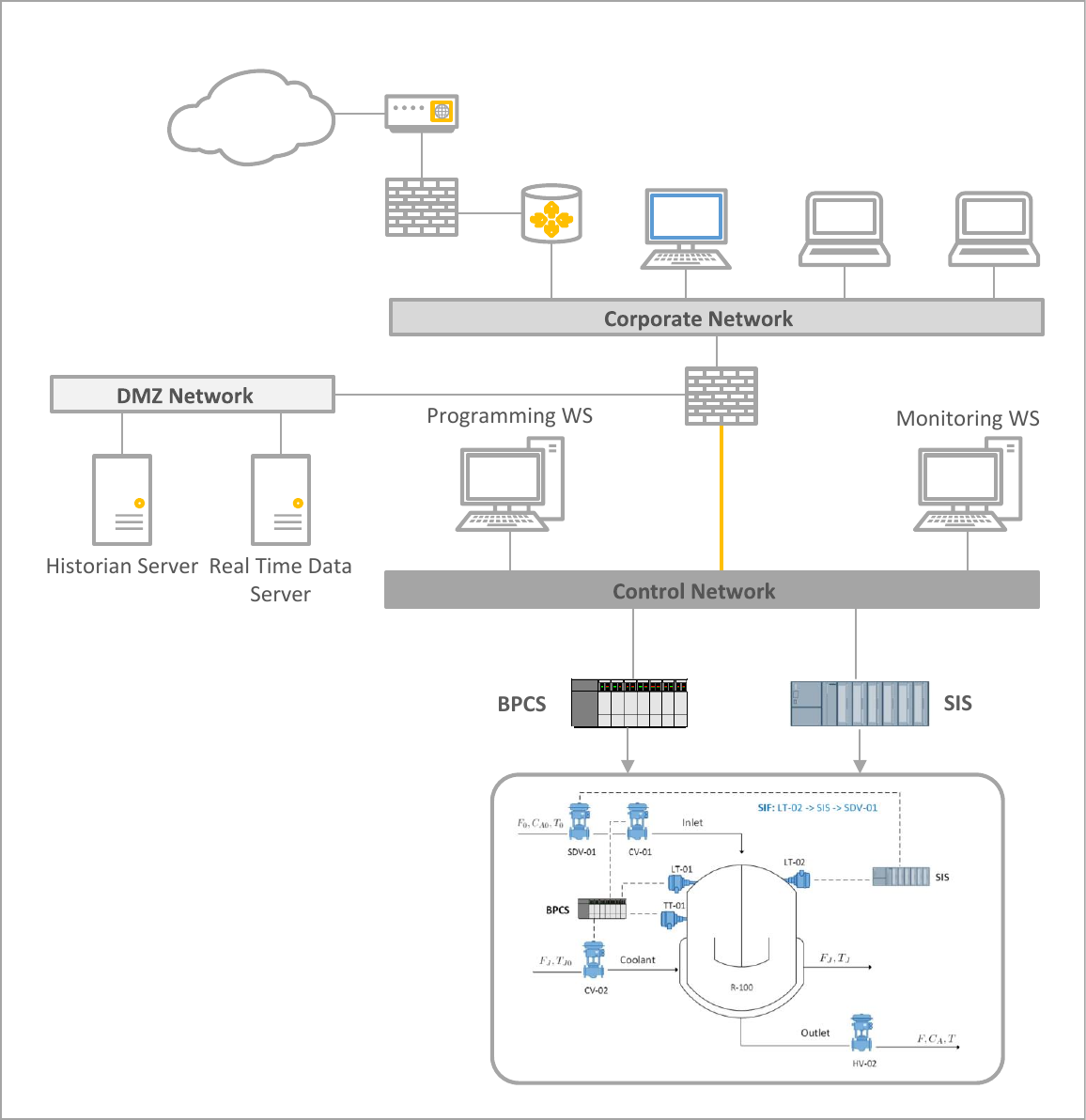}
\caption{CPS architecture for an industrial control system testbed, following NIST 800-82 guidelines. The firewall blocks any direct communication between the corporate network and the control network. Plant floor information is accessible only via the DMZ}
\label{fig:CPS-ARCHI-NIST}
\end{figure}

To compromise the BPCS, we assume the more realistic situation with no insider threat and no direct communication from the corporate network to the control network. In this scenario, the attacker has to detour to compromise the Real Time (RT) server in the DMZ and use it as a pivot to attack the BPCS, either directly or via the monitoring workstation (designated HMI hereafter) that has legitimate communication with the BPCS. We start with the assumption that one of the corporate network PCs that has legitimate access to the RT server is compromised. There are several well-known attack vectors in the IT security domain to achieve such compromise, such as a spam email, a web service vulnerability, or an external malware USB, just to name a few. Figure \ref{fig:Abstract-AT} is an abstract attack tree that summarizes the BPCS compromise paths where the database server compromise is a pre-requisite attack step. In the following, we expand each of the leaf nodes in this abstract attack tree into the corresponding detailed attack trees. More detailed treatment of each attack tree as well as penetration testing could be found in \cite{Tantawy2020}.

\begin{figure}[]
  	\centering
  	\scalebox{0.75}{
  	\scriptsize
  	\begin{tikzpicture}[
  		and/.style={and gate US,thick,draw,fill=gray!40,rotate=90,
  			anchor=east,xshift=-1mm},
  		or/.style={or gate US,thick,draw,fill=gray!40,rotate=90,
  			anchor=east,xshift=-1mm},
  		be/.style={circle,thick,draw,fill=green!60,anchor=north,
  			minimum width=0.7cm},
  		tr/.style={buffer gate US,thick,draw,fill=purple!60,rotate=90,
  			anchor=east,minimum width=0.8cm},
  		label distance=3mm,
  		every label/.style={blue},
  		event/.style={rectangle,thick,draw,fill=white!20,text width=1.75cm,
  			text centered,font=\sffamily,anchor=north},
  		edge from parent/.style={very thick,draw=black!70},
  		edge from parent path={(\tikzparentnode.south) -- ++(0,-1.05cm)
  			-| (\tikzchildnode.north)},
  		level 1/.style={sibling distance=4cm,level distance=1.4cm,
  			growth parent anchor=south,nodes=event},
  		level 2/.style={sibling distance=2cm},
  		level 3/.style={sibling distance=6cm},
  		level 4/.style={sibling distance=3cm}
  		]
  		\node (root) [event] {Reactor overflow Hazard}
  			child {node (g2) {BPCS direct compromise}
  				child {node(g21){BPCS direct attack}}
  				child{node(g22){RT server attack}}
		}
  			child{node(g3){BPCS indirect compromise}
  				child {node(g31){BPCS indirect attack}}
  				child{node(g32){RT server attack}}
  			};
  		\node [or]	at (root.south)	  [label={[label distance=.15cm]-90:$0.033$}]	{};
  		\node [and]	at (g2.south)	[label={[label distance=.15cm]-90:$0.003$}]	{};
  		\node [and]	at (g3.south)	[label={[label distance=.15cm]-90:$0.03$}]	{};
  	\end{tikzpicture}  }	
  	\caption{Abstract attack tree to compromise the BPCS to generate overflow process hazard for the CSTR reactor. Leaf nodes are further expanded in Figures \ref{fig:RTSRV-AT}, \ref{fig:BPCS-AT}, and \ref{fig:HMI-BPCS}.}
  	\label{fig:Abstract-AT}
\end{figure}

Figure \ref{fig:RTSRV-AT} shows the RT server attack tree. The basic idea is to exploit mysql database vulnerabilities via SSH to obtain the Linux server password Hash Dump and possibly crack the password to achieve privilege escalation and gain full control over the RT Server. The probability of success of such an attack depends on several factors including mysql configuration settings to allow brute-force login attack, mysql login password strength, configuration of mysql security monitoring app, and whether SSH is enabled. For the purpose of this case study, we choose this probability arbitrarily as $0.1$. It should be highlighted that the attack tree does not have the sequence semantics to represent a sequence of attack steps. For example, the remote connection step has to be executed before the mysql privileged access in Figure \ref{fig:RTSRV-AT}. We represent this sequence by the \emph{and} gate aggregator, noting that in some other cases the \emph{and} gate may represent simultaneous attack steps. For more information on attack trees and their semantics, the reader is referred to \cite{Mauw2006a}.

\begin{figure}[]
	\centering
	\scalebox{0.75}{
	\scriptsize
	\begin{tikzpicture}[
		and/.style={and gate US,thick,draw,fill=gray!40,rotate=90,
			anchor=east,xshift=-1mm},
		or/.style={or gate US,thick,draw,fill=gray!40,rotate=90,
			anchor=east,xshift=-1mm},
		be/.style={circle,thick,draw,fill=green!60,anchor=north,
			minimum width=0.7cm},
		tr/.style={buffer gate US,thick,draw,fill=purple!60,rotate=90,
			anchor=east,minimum width=0.8cm},
		label distance=3mm,
		every label/.style={blue},
		event/.style={rectangle,thick,draw,fill=white!20,text width=1.5cm,
			text centered,font=\sffamily,anchor=north},
		edge from parent/.style={very thick,draw=black!70},
		edge from parent path={(\tikzparentnode.south) -- ++(0,-1.05cm)
			-| (\tikzchildnode.north)},
		level 1/.style={sibling distance=2.5cm,level distance=1.4cm,
			growth parent anchor=south,nodes=event},
		level 2/.style={sibling distance=3.75cm},
		level 3/.style={sibling distance=2cm},
		level 4/.style={sibling distance=2.5cm}
		]
		\node (root) [event] {RT server compromise}
				child{node(srv-hash){Crack server hashed pass}}
				child{node(db){DB elevated access}
					child{node(remote){Remote connection}
						child{node(rdp){Remote desktop enabled}}
						child{node(ssh-en){SSH enabled}}
					}
					child{node(mysql){mysql privileged access}
						child{node(bf){Brute force login}}
						child{node(pass-cr){Password crack}
							child{node(sql-auth){Bypass mysql authentication}}
							child{node(hash-crack){Crack hashed sql password}}	
						}
					}
				}
				child{node(apparmore){mysql Apparmor disabled}};

		\node [and]	at (root.south)	  [label={[label distance=.15cm]-90:$0.1$}]	{};
		\node [and]	at (db.south)	  [label={[label distance=.15cm]-90:$$}]	{};
		\node[or] at (remote.south) [label={[label distance=.15cm] -90:$$}] {};
		\node[or] at (mysql.south) [label={[label distance=.15cm] -90:$$}] {};
		\node [and]	at (pass-cr.south)	  [label={[label distance=.15cm]-90:$$}]	{};
	\end{tikzpicture}  	}
	\caption{RT Server attack tree. Database vulnerabilities are exploited to gain root access and use the RT server as a pivot to attack the control network. DB elevated access is a prerequisite to crack the server hashed password. Leaf nodes that represent a distinct attack have a weight factor $\gamma_a = 0.024$ that is combined with their success probability. Therefore, Bypass mysql and Crack hashed sql password, Brute force login, SSH enabled, Remote desktop enabled, each has a weight factor combined with their success probability.}
	\label{fig:RTSRV-AT}
\end{figure}

Figure \ref{fig:BPCS-AT} shows the BPCS attack tree, which is divided into two main parts; DoS attack and integrity attack. The DoS attack may not lead to a reactor overflow unless there is a concurrent process disturbance that could not be controlled with the DoS-induced delayed BPCS control response. The probability of such disturbance could be estimated from plant information. The integrity attack injects a low level measurement value for LT-01 to drive the BPCS controller to increase valve CV-01 opening, or directly forces control valve CV-01 to open $100\%$. This will cause a reactor overflow if the SIS is not activated. The injection of the malicious value in the control loop could be accomplished by either gaining access to the controller and overwriting the control program, or more simply sending Modbus packets to the controller with the malicious values. Modbus attack is much easier to launch but requires configuration data to identify the Modbus register address for either LT-01 or CV-01. The probability of BPCS indirect attack is chosen arbitrarily as $0.03$.

\begin{figure*}[]
	\centering
	\begin{minipage}[b]{.65\textwidth}
	\scalebox{0.65}{
		\scriptsize
		\begin{tikzpicture}[
			and/.style={and gate US,thick,draw,fill=gray!40,rotate=90,
				anchor=east,xshift=-1mm},
			or/.style={or gate US,thick,draw,fill=gray!40,rotate=90,
				anchor=east,xshift=-1mm},
			be/.style={circle,thick,draw,fill=green!60,anchor=north,
				minimum width=0.7cm},
			tr/.style={buffer gate US,thick,draw,fill=purple!60,rotate=90,
				anchor=east,minimum width=0.8cm},
			label distance=3mm,
			every label/.style={blue},
			event/.style={rectangle,thick,draw,fill=white!20,text width=2cm,
				text centered,font=\sffamily,anchor=north},
			edge from parent/.style={very thick,draw=black!70},
			edge from parent path={(\tikzparentnode.south) -- ++(0,-1.05cm)
				-| (\tikzchildnode.north)},
			level 1/.style={sibling distance=7.25cm,level distance=1.4cm,
				growth parent anchor=south,nodes=event},
			level 2/.style={sibling distance=5cm},
			level 3/.style={sibling distance=2.5cm},
			level 4/.style={sibling distance=2.5cm}
			]
			\node (root) [event] {BPCS direct attack}
			child{node(dos){DoS attack}
				child{node(cont-dos){Controller under DoS}
					child {node(Modbus){Modbus STOP attack}}
					child{node(con-sd){Controller down}
						child{node (sd) {Shutdown controller}}
						child{node (pass) {Crack password}
							child{node (maxpass) {No max passwd configured}}
							child{node (ssh) {Brute-force SSH login}}
						}
					}
				}
				child{node(dist){Process disturbance}}
			}
			child{node(int){Integrity Attack}
				child {node(prog-ov){Control program overwrite}
					child{node (mal){Run malicious process}}
					child{node(io){I/O access}
						child{node (ioopen){I/O access open}}
						child{node (passcr){Crack password}}
					}
				}
				child{node(modbus-write){Modbus register write}
					child{node (random){Random write hit}}
					child{node (educated){Educated write}
						child{node (config){Get control config}}
						child{node (mod-write){Modbus reg. write}}
					}
				}
			};
			\node [or]	at (root.south)	  [label={[label distance=.15cm]-90:$0.03$}]	{};
			\node [and]	at (dos.south)	  [label={[label distance=.15cm]-90:$$}]	{};
			\node[and] at (con-sd.south) [label={[label distance=.15cm] -90:$$}] {};
			\node[and] at (pass.south) [label={[label distance=.15cm] -90:$$}] {};
			\node [and]	at (prog-ov.south)	  [label={[label distance=.15cm]-90:$$}]	{};
			\node [or]	at (int.south)	  [label={[label distance=.15cm]-90:$$}]	{};
			\node [or]	at (modbus-write.south)	  [label={[label distance=.15cm]-90:$$}]	{};
			\node [or]	at (modbus-write.south)	  [label={[label distance=.15cm]-90:$$}]	{};
			\node [or]	at (io.south)	  [label={[label distance=.15cm]-90:$$}]	{};
			\node [or]	at (cont-dos.south)	  [label={[label distance=.15cm]-90:$$}]	{};
			\node [and]	at (educated.south)	  [label={[label distance=.15cm]-90:$$}]	{};
	\end{tikzpicture}  	}
	\caption{Attack tree for BPCS compromise to generate a reactor overflow hazard. A DoS attack synchronized with a process disturbance or a specially-crafted integrity attack would cause the CSTR to overflow. Leaf nodes that represent a distinct attack have a weight factor $\gamma_a = 0.024$ that is combined with their success probability}
	\label{fig:BPCS-AT}
	\end{minipage} \qquad
	\begin{minipage}[b]{.3\textwidth}
		\scalebox{0.65}{
			\scriptsize
			\begin{tikzpicture}[
				and/.style={and gate US,thick,draw,fill=gray!40,rotate=90,
					anchor=east,xshift=-1mm},
				or/.style={or gate US,thick,draw,fill=gray!40,rotate=90,
					anchor=east,xshift=-1mm},
				be/.style={circle,thick,draw,fill=green!60,anchor=north,
					minimum width=0.7cm},
				tr/.style={buffer gate US,thick,draw,fill=purple!60,rotate=90,
					anchor=east,minimum width=0.8cm},
				label distance=3mm,
				every label/.style={blue},
				event/.style={rectangle,thick,draw,fill=white!20,text width=1.5cm,
					text centered,font=\sffamily,anchor=north},
				edge from parent/.style={very thick,draw=black!70},
				edge from parent path={(\tikzparentnode.south) -- ++(0,-1.05cm)
					-| (\tikzchildnode.north)},
				level 1/.style={sibling distance=4cm,level distance=1.4cm,
					growth parent anchor=south,nodes=event},
				level 2/.style={sibling distance=2cm},
				level 3/.style={sibling distance=2.5cm},
				level 4/.style={sibling distance=2.5cm}
				]
				\node (root) [event] {BPCS indirect attack}
				child{node(hmi){HMI access}
					child{node(pass-leak){Password leak (insider)}}
					child{node(pass-crack){Password crack}
						child{node(rdp-lokcout){No RDP lockout}}
						child{node(bf-rdp){Brute force RDP pass attack}}
					}
				}
				child{node(gui-inj){GUI injection}
					child{node(pr-know){Process knowledge}}
					child{node(gui-know){GUI software knowledge}}
				};	
				\node [or]	at (root.south)	  [label={[label distance=.15cm]-90:$0.3$}]	{};
				\node[or] at (hmi.south) [label={[label distance=.15cm] -90:$$}] {};
				\node [and]	at (pass-crack.south)	  [label={[label distance=.15cm]-90:$$}]	{};
				\node[and] at (gui-inj.south) [label={[label distance=.15cm] -90:$$}] {};
		\end{tikzpicture}}
		\caption{HMI-BPCS indirect attack tree. The compromised HMI is used to embed the attack against the BPCS using the legitimate traffic between the GUI and the BPCS control program. Leaf nodes that represent a distinct attack have a weight factor $\gamma_a = 0.024$ that is combined with their success probability}
		\label{fig:HMI-BPCS}
	\end{minipage}
\end{figure*}

Finally, Figure \ref{fig:HMI-BPCS} shows the attack tree for BPCS attack via the HMI. The attack is launched by remote desktop connection to the HMI and legitimately controlling CV-01 via the GUI. This indirect attack is easier than targeting the BPCS directly as it does not require knowledge about the controller configuration or Modbus register addresses associated with the sensor and valve of the targeted control loop. This is because all the information is already programmed in the GUI software. The probability of BPCS indirect attack is estimated to be $0.3$. Using Figure \ref{fig:Abstract-AT} and the three presented attack trees in Figures \ref{fig:RTSRV-AT}, \ref{fig:BPCS-AT}, and \ref{fig:HMI-BPCS}, the total probability of BPCS attack that leads to an overflow hazard could be estimated by $P[A_B] \approx 0.033$.

It should be highlighted that the assignment of a probability measure to the success of attack actions is subject to debate in the research community, and there is no published agreed-upon data as in the case of reliability failure data. One approach is to use attack databases, such as NIST National Vulnerability Database (NVD) \cite{NIST2016}, to estimate the probability of a cyber attack success based on attributes such as required knowledge level and attack difficulty. However, this approach has the drawback that it does not take into account the specifics of each organization. In this work, we rely on the experience obtained during the penetration testing carried out by the research team in combination with NVD to assign the probability measures. This does not impact the analysis as the presented case study is meant for illustration purposes to explain the design process.

To calculate the probability of BPCS cyber attack leading to a process hazard given a SIS cyber compromise $P[A_{SB}]$, we focus on Modbus attack vectors for both integrity and DoS attacks. Integrity attacks target sensor LT-01 or valve CV-01 as before, either randomly or using leaked Modbus register configuration. DoS attack could be launched by utilizing non-programmed Modbus function code hoping that it would crash the BPCS Modbus master. Figure \ref{fig:SIS-BPCS} summarizes the attack tree, and the probability is chosen arbitrarily as $P[A_{SB}] \approx 0.2813$. To summarize, the desired outcome from the BPCS security lifecycle is $(P[A_B], P[A_{SB}]) = (0.033, 0.2813)$.

\begin{figure}[]
	\centering
	\scalebox{0.65}{
	\scriptsize
	\begin{tikzpicture}[
		and/.style={and gate US,thick,draw,fill=gray!40,rotate=90,
			anchor=east,xshift=-1mm},
		or/.style={or gate US,thick,draw,fill=gray!40,rotate=90,
			anchor=east,xshift=-1mm},
		be/.style={circle,thick,draw,fill=green!60,anchor=north,
			minimum width=0.7cm},
		tr/.style={buffer gate US,thick,draw,fill=purple!60,rotate=90,
			anchor=east,minimum width=0.8cm},
		label distance=3mm,
		every label/.style={blue},
		event/.style={rectangle,thick,draw,fill=white!20,text width=1.5cm,
			text centered,font=\sffamily,anchor=north},
		edge from parent/.style={very thick,draw=black!70},
		edge from parent path={(\tikzparentnode.south) -- ++(0,-1.05cm)
			-| (\tikzchildnode.north)},
		level 1/.style={sibling distance=4.5cm,level distance=1.4cm,
			growth parent anchor=south,nodes=event},
		level 2/.style={sibling distance=1.75cm},
		level 3/.style={sibling distance=2.5cm},
		level 4/.style={sibling distance=2.5cm}
		]
		\node (root) [event] {SIS-BPCS attack}
		child{node(dos){DoS attack}
			child{node(fc){Unsupported Modbus FC}}
			child{node(sis-prog){Modify SIS program}}
			child{node(proc-dist){Process dist.}}
		}
		child{node(int){Integrity attack}
			child{node(rand){Random Modbus injection}}
			child{node(edu){Educated Modbus injection}}
		};	
		\node [or]	at (root.south)	  [label={[label distance=.15cm]-90:$0.28$}]	{};
		\node[and] at (dos.south) [label={[label distance=.15cm] -90:$$}] {};
		\node [or]	at (int.south)	  [label={[label distance=.15cm]-90:$$}]	{};

	\end{tikzpicture}  	}
	\caption{SIS-BPCS attack tree. The SIS is used as a pivot to launch either a DoS attack or a crafted integrity attack that results in reactor overflow. Leaf nodes that represent a distinct attack have a weight factor $\gamma_a = 0.024$ that is combined with their success probability.}
	\label{fig:SIS-BPCS}
\end{figure}

It should be noted that complete attack trees for the given BPCS and CPS architecture could span multiple pages. However, full attack trees may obscure the analysis and will serve no additional insight. Therefore, the simplified attack trees presented here act as a better illustration of the design methodology. For more in-depth treatment of the cyber risk assessment for the presented case study, refer to \cite{Tantawy2020}.
\item SIS Safety Lifecycle - CLOPA: Table \ref{tab:LOPA-CSTR} shows the LOPA sheet for the CSTR overflow hazard identified from the HAZOP, where the initiating event likelihoods and failure probabilities are adopted from \cite{SINTEF2015,CenterforChemicalProcessSafety2015}. The BPCS cyber attack likelihood is calculated as $\lambda_c = 10^{-4}\lambda = 0.01$ /yr, assuming $\lambda = 100$ /yr. Note that human intervention is considered a protection layer assuming there is sufficient time for the operation team to manually isolate the reactor in the field. Some conservative approaches omit any human intervention or safety procedure from the LOPA.

From the LOPA sheet, we extract the event likelihood values to calculate the CLOPA model parameters using equations (\ref{eq:alpha1}) - (\ref{eq:beta}) and (\ref{eq:gamma1}) - (\ref{eq:gamma3}), along with $(P[A_B], P[A_{SB}]) = (0.033, 0.2813)$ from the BPCS security lifecycle. Table \ref{tab:LOPA-PARAMETERS-CSTR} summarizes the parameter values. Substituting in the CLOPA constraint (\ref{eq:LOPA-Symbolic-CPS-COMPONENTS}), we obtain:
\begin{align}     
P[S_p] \leq \frac{1 - 148.68 P[A_S] - 7.6 P[A_{BS}](1 - P[A_S])}{117 (1 - P[A_S]) - 7.6 P[A_{BS}] ( 1 - P[A_S])}
\label{eq:CLOPA-CSTR}
\end{align}
Our objective now is to design a SIF with architecture $\mathcal{A}$ that satisfies (\ref{eq:CLOPA-CSTR}) in order to achieve the required process safety objective as defined by the TMEL in the LOPA analysis. Our initial design for the SIF will comprise a level sensor (LT-02), a logic solver (SIS), and a shutdown valve (SDV-01), as illustrated in Figure \ref{fig:Reactor P&ID}. The SIF will take an independent action upon reactor overflow and will close the inlet shutdown valve. The architecture of the SIF could vary through design iterations to achieve the required safety. As an example, sensors may be duplicated or sometimes triplicated to achieve higher reliability, and the SIS architecture may include redundant CPU modules. We note that for a perfectly-secured SIS $ \left( P[A_S]=P[A_{BS}]=0 \right)$, $P[S_p] \leq 1/117$, or equivalently RRF $\geq 117$. This is the minimum achievable RRF. Since for practical systems there is no zero probability of cyber security attack failures, our SIS design is expected to have an \mbox{RRF $> 117$.}
\begin{table*}[]
\centering
\begin{tabular}{p{4.5cm}p{2.75cm}p{1.5cm}p{1.5cm}p{2cm}p{1.5cm}p{1.5cm}}
	\hline\hline
	Initiating Event & Likelihood $\lambda_i$ (/yr)	& Tank Dike & Safety Procedure & Human Intervention & BPCS ($P[B_p]$) 		& TMEL  \\ \hline\hline
	Inlet flow surge			& $10^{-1}$ 		& $10^{-2}$ & 1 & $10^{-1}$				& $10^{-1}$	& $10^{-6}$ \\ 
	Downstream flow blockage	& $10^{-1}$			& $10^{-2}$ & $10^{-1}$ & $10^{-1}$				& $10^{-1}$	& $10^{-6}$\\ 
	Manual valves misalignment	& $10^{-1}$			& $10^{-2}$ & $10^{-1}$ & $10^{-1}$ 			& $10^{-1}$	& $10^{-6}$\\ 
	BPCS physical Failure				& $10^{-1} (\lambda_b)$			& $10^{-2}$ & 1 & $10^{-1}$			& 1	& $10^{-6}$\\ 
	BPCS attack Failure				& $10^{-2} (\lambda_c)$			& $10^{-2}$ & 1 & $10^{-1}$				& 1	& $10^{-6}$\\ \hline
\end{tabular}
\caption{LOPA sheet for the CSTR overflow hazardous scenario. Numbers in each cell represent the probability of failure of the associated protection layer}
\label{tab:LOPA-CSTR}
\end{table*}

\begin{table}[]
\centering
\begin{tabular}{p{1.2cm} p{2cm} p{4.5cm}}
	\hline\hline
	LOPA Parameter & Value & Source \\ 
	\hline\hline
	$\Sigma_{i=1}^3 \lambda_i$ & 0.3 & LOPA Sheet \\
	$P[\mathcal{L}]$ & 0.001 & LOPA Sheet \\
	$\lambda_b$ & 0.01 & LOPA Sheet \\
	$\lambda_c$ & 0.01 & LOPA Sheet \\
	$P[B_p]$ & 0.01 & LOPA Sheet \\
	$\alpha_1$ & $1.13 \times 10^{-4}$ & CLOPA Parameter-Calculated Eq. (\ref{eq:alpha1}) \\
	$\alpha_2$ & $1.17 \times 10^{-4}$ & CLOPA Parameter-Calculated Eq. (\ref{eq:alpha2})  \\
	$\beta$ & $10^{-6} $ & CLOPA Parameter-Calculated Eq. (\ref{eq:beta})	\\
	$\gamma_1$ & $1.4868 \times 10^{-4}$ & CLOPA Parameter-Calculated Eq. (\ref{eq:gamma1}) \\
	$\gamma_2$ & $7.5785 \times 10^{-6}$ & CLOPA Parameter-Calculated Eq. (\ref{eq:gamma2}) \\
	$\gamma_3$ & $1.1686 \times 10^{-4}$ & CLOPA Parameter-Calculated Eq. (\ref{eq:gamma3}) \\
	\hline
\end{tabular}
\caption{CSTR CLOPA - Calculated parameter values}
\label{tab:LOPA-PARAMETERS-CSTR}
\end{table}

Using the calculated LOPA parameter values, the design region (\ref{eq:design-constraint}) and boundary (\ref{eq:design-boundary}) are defined by:
\begin{align}
P[A_{BS}] \leq 0.132 \left( \frac{1-148.68 P[A_S]}{1-P[A_S]} \right)
\end{align}
where the design boundary is defined when the equality holds. The contour lines for the RRF in (\ref{eq:Contour-Lines}) are defined by:
\begin{align}
P[A_{BS}] = \left(\frac{15.42}{C-1}\right) \left( \frac{(C - 0.008) - (C - 1.27)P[A_S]}{1-P[A_S]} \right)
\label{eq:controu-cstr}
\end{align}
for different values $C$ of the RRF. The design region as well as the contour lines are plotted in Figure \ref{fig:RRF-Contour}. We note that as we approach the design boundary, either by increasing $P[A_S]$ or $P[A_{BS}]$, the RRF rapidly increases such that it is not possible to plot the contour lines in this region in a visible way. The design in this region is very sensitive to input variations (i.e., a very small variation in probabilities will result in a very large change in RRF). Therefore, the design point should be selected as far as possible from the design boundary. To further illustrate the increase in RRF, Figure \ref{fig:RRF-3d} is a 3D plot for the RRF as it varies with both $P[A_S]$ and $P[A_{BS}]$. It should be evident from the 3D plot that for small values of $P[A_S]$, The function gradient is smaller, resulting in a less-sensitive design to probability variations. At larger values of $P[A_S]$ near the design boundary, the RRF increases exponentially with $P[A_{BS}]$. These results could be verified by calculating the gradient of (\ref{eq:LOPA-Symbolic-CPS-COMPONENTS}).
\begin{figure*}[]
	\centering
	\begin{minipage}[b]{.45\textwidth}
		\includegraphics[scale=0.045,trim=12cm 0cm 20cm 4cm,clip]{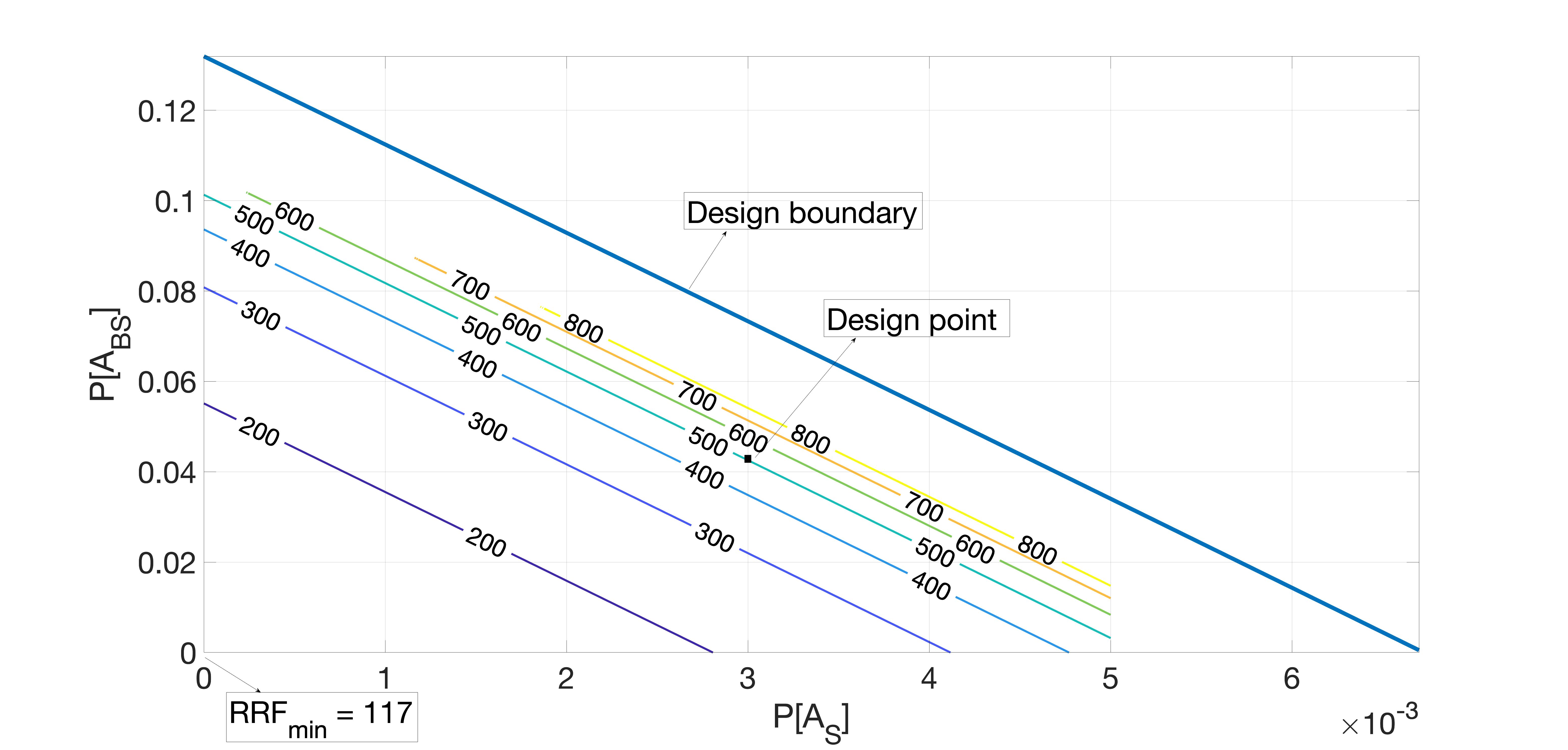}
		\caption{CSTR Case Study: CLOPA design region with contour plot for the Risk Reduction Factor (RRF).}
		\label{fig:RRF-Contour}
	\end{minipage} \qquad
	\begin{minipage}[b]{.45\textwidth}
	\includegraphics[scale=0.06,trim={4cm 2cm 9cm 4cm},clip]{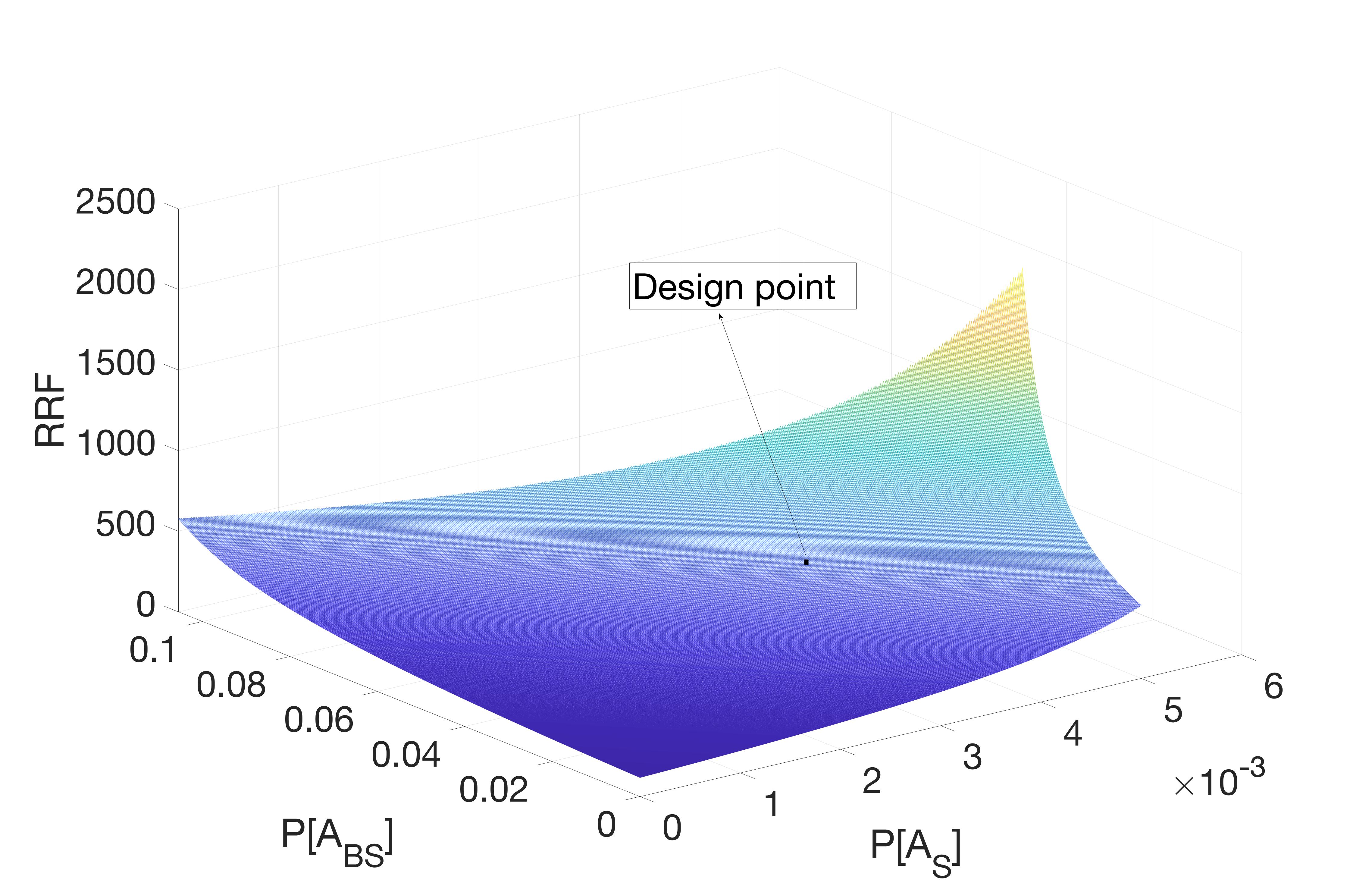}
	\caption{CLOPA RRF as it varies with SIS security failure probabilities. Steepest ascent region to the right should be avoided when selecting the operating point.}
	\label{fig:RRF-3d}
	\end{minipage}
\end{figure*}

To proceed with the design process, we pick the point $P[A_S] = 0.003$ as a reasonable probability value for SIS direct attack failure that is away from the steepest ascent region in Figure \ref{fig:RRF-3d}. We now need to choose a practical value of $P[A_{BS}]$ that results in an achievable target RRF. With the help of Figure \ref{fig:RRF-Contour} and contour lines, $P[A_S]=0.003$ intersects the contour line for RRF = 500 at $P[A_{BS}]=0.0426$. Alternatively, the value of $P[A_{BS}]$ could be obtained from (\ref{eq:controu-cstr}) by setting $C=500$ and $P[A_S]=0.003$. The design point (0.003, 0.0426, 500) is indicated in Figure \ref{fig:RRF-Contour} and \ref{fig:RRF-3d}. The design and verification of the SIF then resumes according to IEC 61511 to develop an architecture $\mathcal{A}$ that satisfies the combined CLOPA requirement: RRF $ \geq 500$, $P[A_S] \leq 0.003$, and $P[A_{BS}] \leq 0.0426$. The detailed design and verification of SIF are outside the scope of the paper (refer to to \cite{IEC61511} for more details). To complete the case study, we will assume the design engineer came up with an architecture $\mathcal{A}$ that was verified using vendor data, resulting in reliability $\text{RRF}_v = 600$, with a design margin from the required $\text{RRF}=500$.

\item SIS Cyber Security Lifecycle: The resulting SIF architecture $\mathcal{A}$ is used to carry out the SIS security lifecycle, similar to the BPCS security risk assessment in step 2 of the design process. As the SIS detailed design and verification is not in the scope of the paper, we will assume for the sake of illustration that the architecture $\mathcal{A}$ results in a cyber system configuration that has a higher probability of SIS security attack failure $P^\prime [A_S]=0.004$ while reducing the BPCS pivot attack failure probability to $P^\prime [A_{BS}]=0.03$ via securing the BPCS-SIS link.

\item Safety Lifecycle - CLOPA: The architecture $\mathcal{A}$ results in $P^\prime [A_S]=0.004$, $P^\prime [A_{BS}]=0.03$, and $\text{RRF}_v = 600$. We need to verify if these values satisfy the CLOPA constraint (\ref{eq:CLOPA-CSTR}). Plugging the probability values results in $P[S_p] \leq 1.54 \times 10^{-3}$, or equivalently $\text{RRF} \geq 643$. As $\text{RRF}_v = 600 < 643$, the architecture has to be modified, either by reducing further the cyber attack failure probabilities, or by increasing the system reliability via fault tolerance techniques. It may take the design engineer multiple iterations until the design achieves the CLOPA constraint. In practice, the iterations do not involve a complete architectural redesign, but rather changing the redundancy scheme or security hardening in order to achieve the design objective. To conclude the case study example, we will assume the design engineer came up with an architecture that preserves the aforementioned probability values while increasing the RRF to 650. This concludes the design process and the system moves to the implementation phase. The case study design values are superimposed on the iterative design process in Figure \ref{fig:Feedback-Design} as an illustration.\hfill $\blacksquare$
\end{enumerate}

\subsection{Classical LOPA Error}
\begin{figure}[]
\centering
\includegraphics[scale=0.06,trim={10.5cm 1.25cm 7cm 6.5cm},clip]{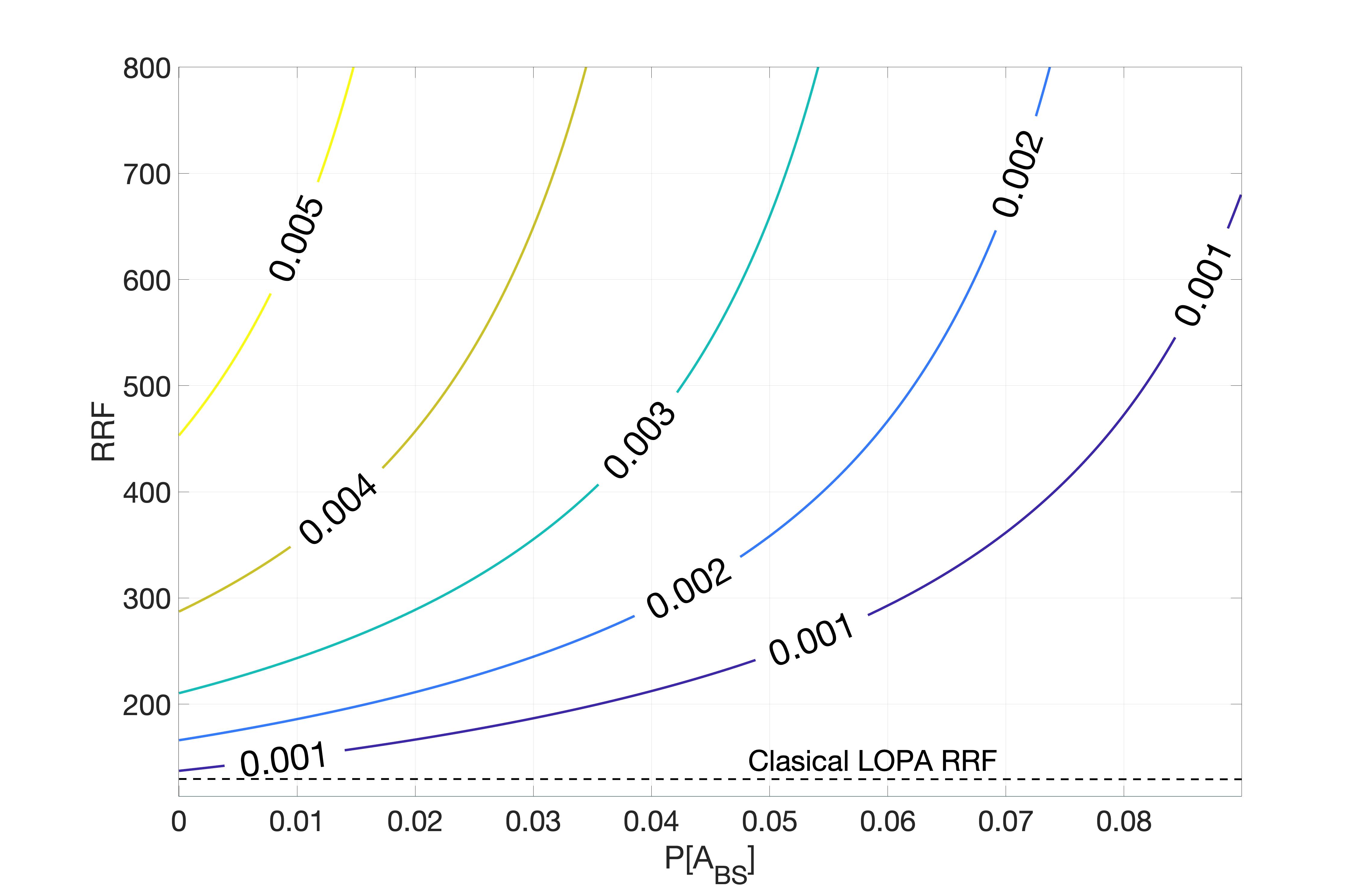}
\caption{Increase of RRF with $P[A_{BS}]$. Each curve corresponds to the fixed value indicated for $P[A_S]$}
\label{fig:Error-Contour}
\end{figure}

Classical LOPA ignores cyber attack probabilities altogether. For the given problem, it results in RRF = 113 as per (\ref{eq:LOPA-SPECIAL}). The minimum CLOPA RRF occurs for a perfectly secured safety system where $P[A_S]=P[A_{BS}]=0$, achieving RRF = 117. Therefore, the minimum error between LOPA and CLOPA RRF estimation is 4. The error gets worse as security failure probabilities increase. For the given design point $P[A_S],P[A_{BS}]=(0.003,0.0426)$, the classical LOPA error is $e_{\text{RRF}}=378$. This is a significant amount of error that results in the design of a less reliable system that will not achieve the target risk level. Figure \ref{fig:Error-Contour} better illustrates the error increase with increasing the security failure probability $P[A_{BS}]$ for different values of $P[A_S]$. For small values of $P[A_S]$, the curves show slow increase in RRF with $P[A_{BS}]$. As $P[A_S]$ increases, the RRF increase becomes exponential. A similar contour figure for fixed $P[A_{BS}]$ values could be generated. The design point for the case study $P[A_S]=0.003$ was chosen as a trade-off between an achievable cyber attack probability value and a moderate rate of increase for the RRF. The 3D plot for the error in RRF vs $P[A_S], P[A_{BS}]$ is identical to Figure \ref{fig:RRF-3d}, except by shifting down the 3D curve by 113, the LOPA RRF value, therefore it is omitted to avoid repetition.

\subsection{Sensitivity Analysis}
Calculating the probability of a security failure is a debatable subject in the research community, especially with lack of statistical data that is available for physical failures. One question that comes to mind is the robustness of the developed CLOPA model to probability variations. We conducted a numerical analysis to calculate the partial derivatives of the RRF with respect to $P[A_S], P[A_{BS}]$.The two partial derivative plots are very similar to Figure \ref{fig:RRF-3d} and omitted for space limitation. For small probability values, the change in the RRF is in the range of 15\% for $10^{-3}$ change in $P[A_S]$. As probabilities increase and we approach the decision boundary, the change in RRF jumps to around 80\% for $10^{-3}$ change and increases exponentially as we get closer to the decision boundary. A similar behavior is exhibited with $P[A_{BS}]$ change (figure omitted for brevity). However, the change in RRF has much lower percentage, ranging from 7\% for small probability values, and increasing to around 37\% as we approach the decision boundary. We highlight the following three key observations: \begin{inparaenum}[(1)]
\item For small cyber failure probability values, the model sensitivity is acceptable since the SIL levels have an order of magnitude ratio, so a small percentage change would likely keep the system requirement in the same SIL category. However, this requires that the probability error is in the range of $10^{-3}$.
\item The model is more sensitive to direct attack failure probabilities than BPCS pivot attacks.
\item We should always try to design our system as far as possible from the decision boundary. The model sensitivity with respect to probability changes increases as we approach the decision boundary.
\end{inparaenum}

\section{Related Work} \label{sec:related-work}
HAZOP has been the dominant risk assessment method for the process industry for over 30 years \cite{Dunjo2010,Crawley2015,Kletz1999}. LOPA has been used in conjunction with HAZOP to design Safety Instrumented Systems (SIS) and specify the Safety Integrity Level (SIL) for each Safety Instrumented Function (SIF) \cite{Dowell1999}. Because of the wide adoption of LOPA by industry due to its systematic approach and quantitative risk assessment capability, LOPA has been included as one of the methods in IEC 61511-3 standard with several illustrating examples \cite{IEC61511}. The LOPA approach has been applied to physical security risk analysis in \cite{Garzia2018}. However, to the best of author's knowledge, there is no research work on integrating security attacks in the LOPA framework for safety instrumented systems design.

There are emergent standardization initiatives to address safety and security coordination in cyber physical systems. IEC 62443-4-1 (Security for industrial automation and control systems - Part 4-1: Secure product development lifecycle requirements) is a standard developed by ISA-99 committee with the purpose to extend existing safety lifecycle at different phases to include security aspects to ensure safe CPS design \cite{IEC2018}. IEC TC65 AHG1 is a recently formed group linked to the same technical committee developing IEC 61508 and IEC 62443 to consider how to bridge functional safety and cyber security for industrial automation systems \cite{Kanamaru2017}. IEC 62859 (Nuclear power plants - Instrumentation and control systems - Requirements for coordinating safety and cyber security) is a standard derived from IEC 62645 for the nuclear power industry to coordinate the design and operation efforts with respect to safety and cyber security \cite{IEC2016}. DO-326 (Airworthiness Security Process Specification) is a standard for the avionics industry that augments existing guidelines for aircraft certification to include the threat of intentional unauthorized electronic interaction to aircraft safety \cite{Torens2020}. A taxonomy of dependable and secure computing is introduced in \cite{Avizienis2004BasicComputing} in order to facilitate the communication among different research communities. The concepts and taxonomy presented are a result of a joint committee on “Fundamental Concepts and Terminology” that was formed by the TC on Fault-Tolerant Computing of the IEEE CS and the IFIP WG 10.4 “Dependable Computing and Fault Tolerance”. A preliminary work on the research in this paper that combines the two research directions stated below is presented in \cite{TantawyICSRS2019}.

\subsection{Lifecycle Integration}
The authors in \cite{Kornecki2013b} use fault tree analysis to combine both safety and security failures in one unified risk assessment framework for the aviation industry. The outcome of the risk assessment is used to define both safety and security requirements. A road-map for cyber safety engineering to increase air traffic management system resilience against cyber attacks is proposed in \cite{Johnson2012}. The V-shaped model to develop embedded software for CPS is augmented with security actions in \cite{Kornecki2010}. The integration of IEC 61508 safety standard and IEC 15408 for IT security is described in \cite{Novak2007a,Novak2008a,Novak2010a} for building automation systems. The author in \cite{Sorby2003} describes in more details the integration of IEC 61508 safety lifecycle and the CORAS approach to identify security risks \cite{Stolen2003}. An approach to align safety and security during different stages of system development lifecycle is proposed in \cite{Hunter2009}. The approach, called Lifecycle Attribute Alignment, ensures compatibility between safety and security controls developed and maintained during the system development lifecycle. HAZOP, a predominantly used method for safety risk assessment in the process industry, is modified in \cite{Winther2001} to include security failures. The authors introduce new guide words, attributes, and modifiers for security components akin to traditional HAZOP limited to safety failures. Failure Mode and Effect Analysis (FMEA) is extended in \cite{Schmittner2014} to include security vulnerabilities, suggesting the name Failure Mode Vulnerability and Effect Analysis (FMVEA). For a survey on the integration of safety and security in CPS, refer to \cite{Lyu2019SafetySystems}.

\subsection{Model-Based Risk Assessment}
Several graphical methods have been used to combine safety and security analysis. Goal Structuring Notation (GSN) is a graphical notation used to model requirements, goals, claims, and evidence of safety arguments \cite{Attwood2011}. The SafSec research project for the avionics industry elaborate on the use of GSN to integrate both safety and security arguments in one representation \cite{Lautieri2007}. A similar approach is used in \cite{Subramanian2016} where authors apply the Non Functional Requirement (NFR) approach to quantitatively assess the safety and security properties of an oil pipeline CPS. NFR is a technique that allows simultaneous safety and security graphical representation and evaluation at the architectural level.

The simplicity and wide adoption of fault and attack trees promoted the research work to merge both modeling tools. The integration of fault trees and attack trees is considered in \cite{NaiFovino2009b} in order to extend traditional risk analysis to include cyber attack risks. A quantitative analysis is proposed by assigning probabilities to tree events. Similarly, fault tree analysis is used in \cite{Kornecki2013} to analyze safety/security risks in aviation software. In \cite{Steiner2013}, the authors extend Component Fault Trees (CFT) to contain both safety and security events. Both qualitative and quantitative analysis is performed to assess the overall risk. The quantitative analysis is enabled by assigning probabilities to safety events and categorical rating (low, medium, high) for security events. The authors in \cite{Kumar2017} translate the combined fault-attack tree into stochastic time automata to enable quantitative risk analysis. The use of Bow-tie diagrams and analysis in place of fault trees is reported in \cite{Abdo2018b}, where it is integrated with attack trees for combined safety-security risk assessment.

Given the limited semantics of fault trees, Boolean logic Driven Markov Process (BDMP) graphical formalism introduced in \cite{Bouissou2003} has been used to integrate safety and security events. The approach integrates fault trees with Markov process at the leaf nodes level and associates a mean time to success (MTTS) for security events and a mean time to failure (MTTF) for safety events. This allows both a qualitative and a quantitative risk assessment for the given system. The formalism also enables the modeling of detection and response mechanisms without a need for model change. The work in \cite{Kriaa2014} applies BDMP formalism to a pipeline case study, illustrating different types of safety-security inter-dependencies. In \cite{Kriaa2012a}, Stuxnet attack is modeled using BDMP and a quantitative risk analysis is carried out on the industrial control system.

Petri nets have also been proposed to overcome the limitations of fault trees. A formalism for safety analysis named State/Event Fault Trees (SEFTs) is reported in \cite{Kaiser2007}. In this formalism, both deterministic state machines and Markov chains are combined, while keeping the visualisation of causal chains known from fault trees. This formalism is extended in \cite{Roth2013} to include an attacker model to deal with both safety and security. Similarly, stochastic Petri nets have been used in \cite{Mitchell2013} to model the impact of intrusion detection and response on CPS reliability, and in \cite{Ten2008} to assess the vulnerabilities in SCADA systems. Bayesian belief networks are also considered as one of the model-based approaches. In \cite{Kornecki2013a}, a Bayesian Belief Network is used to assess the combined safety and security risk for an oil pipeline example.

The Unified Modeling Language (UML) commonly used in software engineering has also been used for safety and security risk assessment. Misuse cases for UML diagrams have been used to define safety requirements in \cite{Sindre2007} and security requirements in \cite{Sindre2005}, independently. A combined process for Harm Assessment of Safety and Security has been proposed in \cite{Raspotnig2012} based on both UML and HAZOP studies. UMLsafe \cite{Jurjens2003} and UMLsec \cite{Jurjens2002} are two UML extensions that enable modeling of safety and security requirements, respectively. The combined UMLsafe/UMLsec is proposed in \cite{Jurjens2003a} for safety-security co-development. SysML-sec, a SysML-based model driven engineering environment, is used in \cite{Pedroza2011} for the formal verification of safety and security properties.

System Theoretic Process Analysis (STPA) was developed as a new hazard analysis technique to evaluate the safety of a system \cite{Thomas2013}. The authors in \cite{Friedberg2017} extend the STPA to include system security aspects in the analysis. The expanded approach is named STPA-SafeSec and demonstrated on a use case in the power grid domain. The System Theoretical Accident Model and Process (STAMP) is applied to the Stuxnet attack in \cite{Nourian2018AStuxnet}, showing that the attack could have been avoided if STAMP was applied during design time.

\section{Conclusion} \label{sec:conclusion}
Classical safety assessment methods do not take into account failures due to cyber attacks. In this paper, we showed quantitatively that overlooking security failures could bias the risk assessment, resulting in under-designed protective systems. In addition, the design of safety and security subsystems for complex engineering systems cannot be carried out independently, given their strong coupling as demonstrated in this paper. Although the design becomes more complicated when considering cyber attacks, the development of new software tools or the modification of existing industrial tools could automate the process.

In this work, we considered the control system (BPCS) design as given, following common industrial practice. Joint optimization of both BPCS and SIS designs, from both safety and security perspectives, is a potential extension for the presented work. Also, the presented integrated lifecycle relies in part on designer's experience to make design decisions to achieve the system requirements. Optimal system design that captures possible safety and security design choices with associated financial cost could provide a better quantitative approach to find the optimal system operating point rather than relying on design heuristics. Furthermore, the integration of both the safety and security lifecycles into model-based design toolchains is crucial for adoption by industry.

Finally, the work presented in this paper discusses the impact of cyber security failure on system safety. A closely-related problem is how safety failures could impact cyber security. There is not much work in this direction, perhaps because the focus in cyber physical systems is always on safety, considering the security of the cyber system as a secondary issue. Nevertheless, this is an important problem. On one hand, a simple safety failure may be injected to cause a security compromise that may be exploited to produce a higher security compromise that could lead to a greater safety hazard. On the other hand, both directions, i.e., Safety $\rightarrow$ Security and Security $\rightarrow$ Safety, are closely related and interacting, and therefore optimizing a cyber physical system performance with respect to safety/security or both cannot be fully achieved without understanding the two types of interactions.

\appendices

\section{Proof of Lemma \ref{lemma:ATTACK-AGGREGATION}} \label{app:lemma-proof}
\begin{proof}
The aggregate likelihood of all attacks to cause a hazard taking into account BPCS and SIS protection could be approximated by (neglecting higher order probability terms):
\begin{align}
 \Lambda = \lambda \sum_{a \in \mathcal{A}_r} \alpha_a P_a[S,B]
\end{align}
Using (\ref{eq:Joint-Prob}) to expand the joint probability:
\begin{align}
\Lambda = \lambda \sum_{a \in \mathcal{A}_r}  \alpha_a \left( \eta_1 + \eta_2 P[B^a_c] + \eta_3 P[S_c] P[B^a_c | S_c] \right)
\label{eq:Sum-Attacks}
\end{align}
where $P[B^a_c]$ represents the probability of BPCS security failure with respect to attack $a$, and $\eta_1, \eta_2, \eta_3$ are probability terms not dependent on the attack $a$. Expanding:
\begin{align}
\Lambda &= \lambda \left( \sum_{a \in \mathcal{A}_r} \alpha_a \right) \times \\
& \left( \eta_1 + \eta_2 \sum_{a \in \mathcal{A}_r} \gamma_a P[B^a_c] + \eta_3 \sum_{a \in \mathcal{A}_r} \gamma_a P[S_c] P[B^a_c | S_c] \right)
\end{align}
where $\gamma_a = \alpha_a / \sum_{a \in \mathcal{A}_r} \alpha_a$. Ignoring higher order probabilities:
\begin{align}
\Lambda &\approx \lambda \left( \sum_{a \in \mathcal{A}_r} \alpha_a \right) \times \\
& \left( \eta_1 + \eta_2 P \left[ \sum_{a \in \mathcal{A}_r} \gamma_a B^a_c \right] + \eta_3 P \left[ S_c, \sum_{a \in \mathcal{A}_r} \gamma_a B^a_c \right] \right)
\label{eq:Equivalent-Attack}
\end{align}
Comparing (\ref{eq:Sum-Attacks}) and (\ref{eq:Equivalent-Attack}), the second and third terms in (\ref{eq:Equivalent-Attack}) represent an equivalent BPCS with a combined attack vector $\mathcal{A}_r$, where each attack $a$ is weighted by $\gamma_a$. In addition, the likelihood of this combined attack vector is $\lambda \left( \sum_{a \in \mathcal{A}_r} \alpha_a \right)$.
\end{proof}

\section*{Source Code}
The source code for the CLOPA in the form of Matlab m files to regenerate the research results including the case study are located at https://github.com/Ashraf-Tantawy/CLOPA.git

\section*{Acknowledgment}
This research was made possible by NPRP 9-005-1-002 grant from the Qatar National Research Fund (a member of The Qatar Foundation). The statements made herein are solely the responsibility of the authors.

\begin{table*}[]
\centering
\begin{tabular}{p{1.5cm} p{6cm} p{3.5cm} p{4cm}}
	\hline\hline
	Symbol  &   Description &   Type & Calculation Method/ Data Source  \\
	\hline\hline
	$\lambda_i$   &   Initiating event $i$ likelihood (/yr)&   Parameter & Reliability data   \\
	$\lambda_p$ & BPCS physical failure event likelihood (/yr) & Parameter  & Reliability data\\
	$\lambda_a$ & BPCS cyber attack $a$ likelihood (/yr) & Parameter  & Refer to Section \ref{sec:symantec-attack}\\
	$\lambda_c$ & BPCS semantically-related attacks likelihood (/yr) & Parameter  & Refer to Section \ref{sec:symantec-attack}, Lemma \ref{lemma:ATTACK-AGGREGATION}\\
	$\lambda$ & BPCS cyber attack likelihood for all attacks & Parameter  & Statistical attack data \\
	$\alpha_a$ & Probability of selecting attack $a$ by the attacker & Parameter  & Attacker profile model \\
	$\gamma_a$ & Weight factor for the attacks for the equivalent BPCS & Parameter  & Refer to Lemma \ref{lemma:ATTACK-AGGREGATION} \\
	$P[\mathcal{L}_i]$ & Probability of failure of all protection layers for initiating event $i$ & Parameter & Reliability data \\
	TMEL & Target Mitigated Event Likelihood & Parameter & Determined by the corporate policy \\
	$P[B_c]$    &  Probability of BPCS security failure & Intermediate design variable & BPCS security risk assessment  \\
	$P[B_p]$    &  Probability of BPCS physical failure & Parameter & Reliability data  \\
	$P[A_B]$    & Probability of BPCS direct security failure & Parameter & BPCS security risk assessment  \\
	$P[A_{SB}]$    &  Probability of BPCS SIS-pivot security failure & Parameter & BPCS security risk assessment \\
	$P[S_c]$    &  Probability of SIS security failure & Intermediate design variable & SIS security risk assessment \\
	$P[S_p]$    &  Probability of SIS physical failure & Design variable  & SIS security risk assessment \\
	$P[A_S]$    & Probability of SIS direct security failure & Design variable & SIS security risk assessment \\
	$P[A_{BS}]$    &  Probability of SIS BPCS-pivot security failure & Design variable & SIS security risk assessment  \\
	$P[S_c,B_c]$ & Probability of simultaneous SIS and BPCS security failure & Intermediate design variable & BPCS \& SIS security risk assessment \\ 
	$\alpha_1 - \alpha_2$ & - & Auxiliary parameters & Eq. (\ref{eq:alpha1}), (\ref{eq:alpha2})  \\
	$\gamma_1 - \gamma_3$ & - & Auxiliary parameters  & Eq. (\ref{eq:gamma1}) to (\ref{eq:gamma3}) \\ 
	$\zeta_1 - \zeta_3$ & - & Auxiliary parameters  & Eq. (\ref{eq:zeta1}) to (\ref{eq:zeta3}) \\ 
	$\beta$ & - & Auxiliary parameters & Eq. (\ref{eq:beta})  \\
	\hline
\end{tabular}
\caption{CLOPA model parameters. Variables designated as "Design variable" are with respect to CLOPA, but could be a design variable of another assessment, such as $P[A_B]$, derived from BPCS security risk assessment. Variables designated as "Intermediate design variables" could be expressed in terms of design variables.}
\label{tab:Model-Parameters}
\end{table*}

\bibliographystyle{ieeetr}
\bibliography{references}

\flushend
\end{document}